\documentclass{aa}  

\usepackage{graphicx}
\usepackage{xcolor}
\usepackage{amsmath}
\usepackage{acronym}

\usepackage{txfonts}
\begin{document} 

\title{High latitude emission from structured jet of $\gamma$-Ray Bursts observed off-axis}
\titlerunning{Structured jet}

\author{S. Ascenzi \inst{1, 2}\thanks{E--mail: stefano.ascenzi@inaf.it},
G. Oganesyan \inst{2,3,4}, O. S. Salafia \inst{1, 5}, M. Branchesi \inst{2,3,4}, G. Ghirlanda \inst{1}, S. Dall'Osso \inst{2, 3}}

\institute{$^1$  INAF -- Osservatorio Astronomico di Brera, Via Bianchi 46, I--23807 Merate, Italy \\
$^2$ Gran Sasso Science Institute, Viale F. Crispi 7, I-67100, L’Aquila (AQ), Italy\\
$^3$ INFN - Laboratori Nazionali del Gran Sasso, I-67100, L’Aquila (AQ), Italy\\
$^4$ INAF - Osservatorio Astronomico d’Abruzzo, Via M. Maggini snc, I-64100 Teramo, Italy\\
$^5$ INFN – Sezione di Milano-Bicocca, Piazza della Scienza 3, 20126 Milano, Italy}

\authorrunning{Ascenzi, Oganesyan, Salafia, Branchesi, Ghirlanda \& Dall'Osso}

\newcommand{\sa}[1]{{\textcolor{blue}{\texttt{SA: #1}}}}
\newcommand{\gor}[1]{{\textcolor{cyan}{\texttt{gor: #1}}}}
\newcommand{\om}[1]{{\textcolor{orange}{\texttt{Om: #1}}}}
\newcommand{\gh}[1]{{\textcolor{red}{\texttt{GG: #1}}}}

\abstract{

The X--ray emission of \acp{GRB} is often characterized by an initial steep decay, followed by a nearly constant emission phase (so called `plateau') which can extend up to thousands of seconds. While the steep decay is usually interpreted as the tail of the
prompt $\gamma$--ray flash, the long-lasting plateau is commonly associated to the emission from the external shock sustained by energy injection from a long lasting central engine.  A recent study proposed an alternative interpretation, ascribing both the steep decay and the plateau to high-latitude emission (HLE) from a `structured jet' whose energy and bulk Lorentz factor depend on the angular distance from the jet symmetry axis. 
In this work we expand over this idea and explore more realistic conditions: (a) the finite duration of the prompt emission, (b) the angular dependence of the optical depth and (c) the lightcurve dependence on the observer viewing angle. We find that, when viewed highly off-axis, the structured jet HLE lightcurve is smoothly decaying with no clear distinction between the steep and flat phase, as opposed to the on-axis case. 
For a realistic choice of physical parameters, 
the effects of a latitude-dependent Thomson opacity and finite duration of the emission have a marginal effect on the overall lightcurve evolution.
We discuss the possible HLE of GW170817, showing that the emission would have faded away long before the first \emph{Swift}-XRT observations. Finally, we discuss the prospects for the detection of HLE from off-axis GRBs by present and future wide-field X-ray telescopes and X-ray surveys, such as eROSITA and the mission concept THESEUS.
}

\keywords{gamma--ray burst: general}

\maketitle
%\onecolumn
%
%-------------------------------------------------------------------

%\begin{acronym}
\acrodef{HLE}{high-latitude emission}
\acrodef{GRB}[GRB]{$\gamma$-ray burst}
\acrodef{EATR}[EATR]{Equal Arrival Time Ring}
\acrodef{CoE}[CoE]{center-of-explosion}
\acrodef{GW}[GW]{gravitational-waves}
\acrodef{EM}[EM]{electromagnetic}
\acrodef{SBPL}[SBPL]{smoothly broken power law}
\acrodef{BNS}[BNS]{binary neutron star}
\acrodef{LVC}[LVC]{LIGO and Virgo Scientific Collaborations}
\acrodef{ET}[ET]{Einstein Telescope}
\acrodef{SXI}[SXI]{Soft X-ray Imager}
%\end{acronym}

\section{Introduction}

The observed prompt emission in the keV--MeV range and the multi-wavelength afterglow clearly indicate that \acp{GRB} originate in highly relativistic, collimated jets (see e.g.~\citealt{Piran2004} and \citealt{KumarZhang2015} for reviews). 

At lower energies, their early lightcurve shows a complex behaviour (e.g. \citealt{Tagliaferri2005, Nousek2006, Zhang2006, Brien2006, Liang2007, Willingale2007}): frequently, the optical and X--ray emission present different temporal decays and, more often, the X-ray lightcurve presents a steep decay followed by a plateau, while the optical emission is consistent with a smooth decay 
(e.g. \citealt{Fan2006,Panaitescu2008,Ghisellini2007,Liang2008, Panaitescu2011,Oates2011, DePasquale2011, Zaninoni2013, Melandri2014, Li2015}). The origin of the plateau is a long-standing problem of \ac{GRB} physics, whose standard interpretation invokes the presence of a central engine with a long lasting activity, such as a millisecond magnetar formed after a binary neutron star merger or a stellar collapse \citep{Dai1998a,Dai1998b,Dai2004,ZhaMes2001, Yu2010,Metzger2011, DallOsso2011}. 

Both the internal and external jet's kinetic energy dissipation mechanisms are expected to contribute to the diversity of the observed X-ray and optical emission (\emph{e.g.} see \citealt{Li2015}). A model including internal and external jet dissipation is required to explain the chromatic nature of the X-ray and optical lightcurves (e.g. plateau in X-rays and normal decay in optical bands, see e.g.
\citealt{Fan2006, Paper1}) the different post-plateau phases (including rarely observed fast decay, see e.g. \citealt{Sarin2020}), and the claimed correlations between the observed plateau/prompt emission properties (e.g. \citealt{Dainotti2008, Margutti2013, Izzo2015, Tang2019}).

 \citet{Paper1} and \citet{BeniaminiDuque2020} showed that a structured jet could provide an alternative explanation for the origin of the X-ray plateau in \ac{GRB}. A structured jet is an outflow where velocity and energy vary (decreasing, in general) with the angle measured from the jet axis, as opposite to a top-hat jet, in which these quantities are uniform within the jet opening angle. \citealt{Paper1} (hereafter, O20) assumes an instantaneous \textit{prompt emission} pulse originated from a structured jet, showing that the regions outside the jet core can produce a long lasting X-ray plateau for an on-axis observer. 

This effect is produced by the \ac{HLE} \emph{i.e.} the emission produced at large angles with respect to the jet axis, where the difference in the time of flight of photons (with respect to photons emitted from the axis) is high enough\footnote{Namely larger than the temporal resolution of the detector.} to shape the lightcurve in the observer frame.
While \ac{HLE} has been already considered to explain the tails of the prompt emission pulses \citep{Fenimore1996}, the X-ray steep-decay phase \citep{Kumar2000, Zhang2006,Liang2007}, and the tails of the X-ray flares \citep{Wei2015}, its role in producing the X-ray plateau is a novelty, introduced for the first time in the O20 model\footnote{A flattening in the decay of the \ac{HLE} lightcurve in the case of a structured jet have been previously pointed out by \citet{Dyks2005}}. A particular feature of this model is that the duration of plateau phase increases with the distance from the central engine where the emission happens. For appropriate combination of parameters O20 showed that plateaus as long as few $10^4\,\rm s$ can be produced.

Alternatively, \citealt{BeniaminiDuque2020} showed that the plateau emission can be interpreted as the external shock (\textit{afterglow}) emission of a structured jet observed slightly off-axis. 
It is worth noticing that both \citet{BeniaminiDuque2020} and O20 models have the great advantage of being agnostic on the nature of the \ac{GRB} central engine.

The idea that GRB jets are structured is  supported by several theoretical studies, which investigated how the jet propagates through the envelope of the massive stellar progenitor in long \acp{GRB} or through the ejecta resulting from the merger of the compact binary progenitor in short \acp{GRB} (e.g.~\citealt{Aloy2000, RR2002, Matzner2003, Zhang2003, Lazzati2005, Aloy2005,Morsony2007, Bromberg2011,Lazzati2019,Xie2019, Om2019}). These studies have shown that the jet-envelope/ejecta interaction leaves an imprint on the jet, leading to the formation of a wide angular structure. 

The recent joint discovery of the \ac{GW} signal from the \ac{BNS} merger GW170817, of the accompaining faint short GRB~170817A \citep{Abbott2017a,Abbott2017b}, and in particular, of its multi-wavelength long lasting emission, has provided a strong support to the structured jet scenario (e.g.~\citealt{Lazzati2018,Davanzo2018,Mooley2018,Resmi2018,Lyman2018,Margutti2018,Troja2018,Ghirlanda2018, Lamb2019,Hajela2019, Salafia2019,Ioka2019}).  

Here we aim to generalize the O20 model by relaxing some of its basic assumptions, which are:
i) the coincidence of the line of sight with the jet axis (on--axis observer), ii) the infinitesimal duration of the energy release by the emitting surface and iii) the negligible opacity of the emitting region. 
We generalize the model considering: i) a generic viewing angle with respect to the jet axis, ii) a finite duration in the energy release and iii) a non negligible opacity and its angular dependence along the emitting surface.

Moreover, the off-axis view of the \ac{GRB} prompt \ac{HLE} gives origin to a new promising class of X-ray counterparts of \ac{GW} from compact binary mergers, which add up to the already explored orphan afterglows from the external shock, cocoon emission (see \citealt{Lazzati2017b,Lazzati2017a}), and spindown--powered transients \citep{YuZhangGao2013, MePi2014, SiegelCiolfi2016a, SiegelCiolfi2016b} (for more details on electromagnetic counterparts from \acp{GW} see \citealt{Nakar2019}). 

The \acp{HLE} can become a detectable target in the near future, thanks to the advent of sensitive large field-of-view surveys, such as eROSITA \citep{eROSITA2012}, Einstein Probe \citep{EinsteinProbe2015}, eXTP \citep{Hernanz2018}, the mission concept THESEUS-\ac{SXI} \citep{Theseus2018} and SVOM-ECLAIRs \citep{SVOM2016} in the X-rays.
Thus, its modelling is of great importance in order to plan observational strategies for the follow-up of \acp{GW} events, detected by the current and next generation \ac{GW} detectors \citep{ETdesignstudy, ET2019, Reitze2019},  and to optimize the operations and development of wide-field X-ray instruments \citep{Theseus2018, Hernanz2018}. 

The paper is organized as follows: in Section \ref{sec:instantaneous} we describe the formalism adopted for the computation of the \ac{HLE} from a structured jet for any observer's viewing angle: we first develop the simplest case of an instantaneous emission (\S 2.1) and then generalize to the case of finite duration of the emission (\S 2.2). In Section \ref{sec:opacity} we derive the transparency condition to Thomson scattering as a function of the jet's angular coordinate and discuss its impact on the observed X-ray lightcurve. In Section \ref{sec:results} we show our resulting X-ray lightcurves and discuss the detectability prospect of off-axis events by THESEUS and  eROSITA. In this section, we address also the case of GW/GRB 170817. 
We discuss our findings in Section \ref{sec:summary}. 

%--------------------------------------------------------------------
\section{Method}
\label{sec:instantaneous}
In this section we outline the formalism for the computation of the HLE for an observer at any viewing angle (i.e.~relaxing the assumption of an on-axis observer adopted in O20). We first consider (similary to O20) the case of an instantaneous emission and then extend the treatment to the more realistic case of a finite duration pulse. Finally, we include also the effect of a non negligible opacity.

\subsection{Instantaneous Emission}

Following O20 we start by assuming that all the energy is released instantaneously at the emission time $t_{\rm em} = R_0/c/\sqrt{1-1/\Gamma^2_c}$.
Here $R_{0}$ is the distance from the central engine along the jet axis ($\theta = 0$)
at the time when
the energy is released and $\Gamma_{\rm c}$ is the bulk Lorentz factor of the core of the structured jet.

We consider two possible jet structures, which describe how the energy and the bulk Lorentz factor change as a function of the angle from the jet axis. These have been first proposed in models for the afterglow emission (power-law structured jet \citep{Rossi2002}; Gaussian structured jet \citep{Zhang2002}). The power-law structure is defined as
\begin{equation}
    \begin{cases}
        \epsilon(\theta) = \Bigl[1+\Bigl(\frac{\theta}{\theta_E}\Bigr)^k\Bigr]^{-1}\\
        \Gamma(\theta) = 1 + (\Gamma_c -1)\Bigl[1 + \Bigl(\frac{\theta}{\theta_c}\Bigr)^k\Bigr]^{-1}
    \end{cases}
    \label{eq:structPL}
\end{equation}
The Gaussian structure is defined as
\begin{equation}
    \begin{cases}
    \epsilon(\theta) = e^{-(\theta/\theta_E)^2}\\
    \Gamma(\theta) = 1+(\Gamma_c -1)e^{-(\theta/\theta_c)^2}
    \end{cases}
    \label{eq:structGaussian}
\end{equation}
where $\theta_{\rm E}$ and $\theta_{c}$ are the core opening angles of the (normalized) energy and Lorentz factor structure, respectively, and $\Gamma_{\rm c}$ is the bulk Lorentz factor of the core\footnote{
Differently from O20, we choose here a smooth function for the power-law structure, which ensures the stability of our algorithm for the \ac{HLE} computation.}.
\begin{figure}
    \centering
    \includegraphics[width=0.46\textwidth]{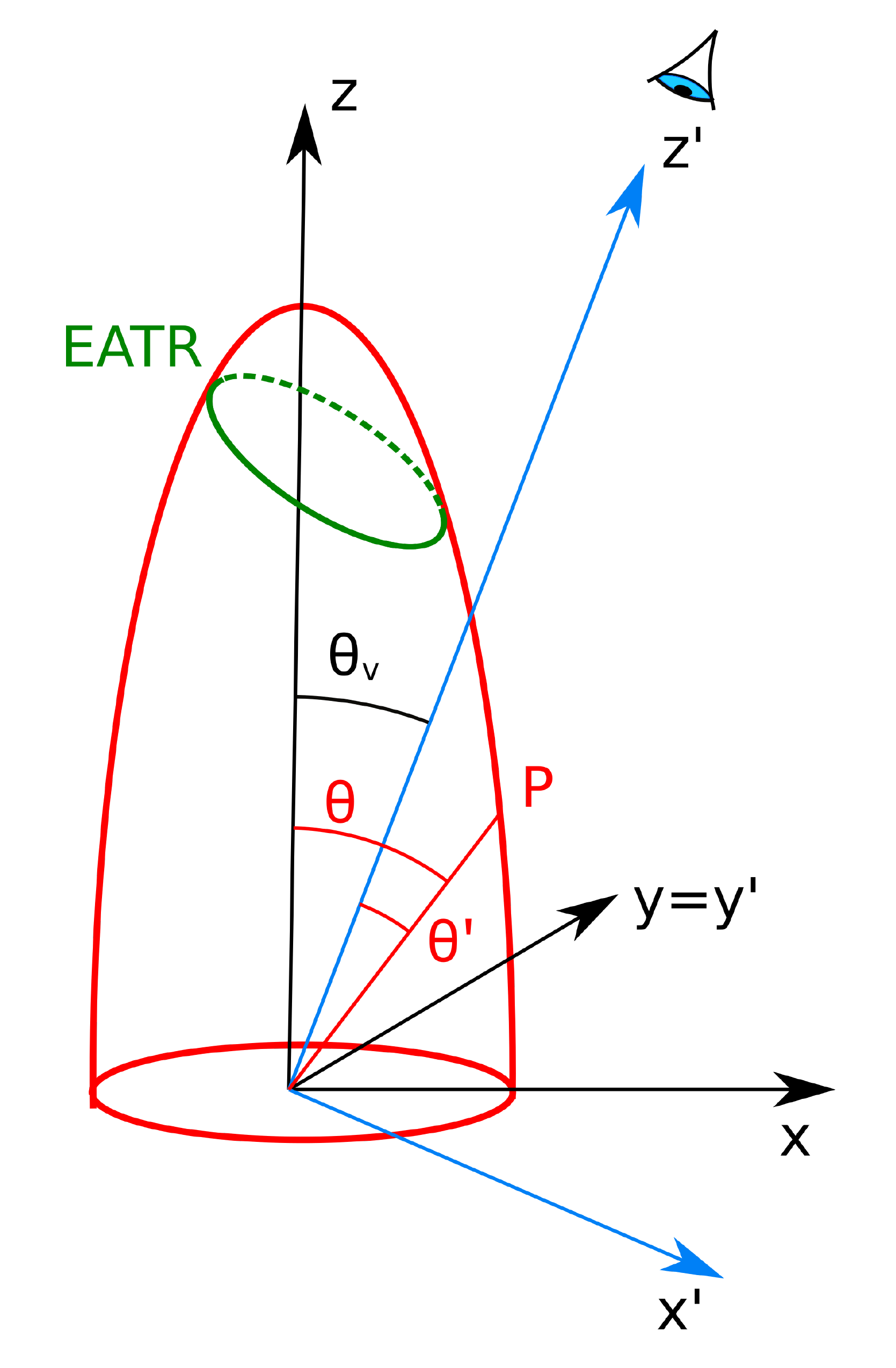}
    \caption{Pictorial representation of the system under study. The emitting surface is shown in red, the reference frame $\mathcal{K}$ is in black and the frame $\mathcal{K}'$, with its $z'$ axis aligned with the line of sight, is represented in blue. The green circle shows an example of EATR.}
    \label{fig:sketch}
\end{figure}
In order to explore how the HLE from the structured jet \ac{GRB} prompt emission appears under different viewing angles, we consider
two different Cartesian coordinate systems $\mathcal{K}$ and $\mathcal{K}'$ represented in Fig. \ref{fig:sketch} (black and blue axes,  respectively). The $z$--axis of the $\mathcal{K}$ system is aligned with the jet's symmetry axis, while the $\mathcal{K}'$ system is rotated by an angle $\theta_v$ (the viewing angle) around the $y$ axis such that the axis $z'$ coincides with the line of sight\footnote{The observer is assumed to lie in the x-z plane, without lack of generality thanks to the assumed axial symmetry.}. Both systems are centered on and at rest with respect to the central engine, namely they are \ac{CoE} frames.

At a given observer time $t_{\rm obs}$, the observer receives radiation emitted by those points of the emitting surface sharing the same value of $z'$. Such points define a curve on the emitting surface called \ac{EATR}, which is described as (see Appendix \ref{sec:EATReq}):
\begin{equation}
    R(\theta)\bigl(\sin\theta_v\sin\theta\cos\phi + \cos\theta_v\cos\theta\bigr) = -c(t_{\rm obs} - t_{\rm em}). 
\end{equation}
where $R(\theta) \equiv \beta(\theta)c t_{\rm em}$ is the radius at given $\theta$ that defines the surface of emission, $\beta(\theta)$ the jet velocity in unit of $c$, and $t_{\rm em}$ denotes the time at which the emission occurs (all the times are measured starting from the jet launching time). 

The flux at a given $t_{\rm obs}$ is  obtained by integrating the flux element along the \ac{EATR} corresponding to a given $t_{\rm obs}$, taking into account also the Doppler factor of the emitting element. 
For an off-axis observer, the break of the symmetry introduces in the Doppler factor a dependence on the angle $\phi$, such that it writes:
\begin{equation}
    D(\theta, \phi) = 
    \frac{1}{\Gamma(\theta)[1-\beta(\theta)(\cos\theta\cos\theta_v + \sin\theta\sin\theta_v\cos\phi)]},
    \label{eq:dopplerfactor}
\end{equation}
where $\cos\theta \cos \theta_v + \sin \theta \sin \theta_v \cos \phi = \cos \theta'$ and $\theta'(\theta, \phi)$ is the polar angle of the point P in the $\mathcal{K}'$ (see Fig. \ref{fig:sketch} and Appendix \ref{sec:EATReq}). It is worth noticing that, while in the on-axis case the Doppler factor depends only on $\theta$ (e.g. see O20), for an off-axis observer, the break of the symmetry introduces a further  dependence on the angle $\phi$.

The flux density thus can be written as:

\begin{align}
    &F_\nu (t_{\rm obs}) = \frac{1}{d^2_L(1+z)^3} \times\notag \\ &\oint_{EATR} \frac{{\eta'}_{\nu' }[\nu(1+z)/D(\theta, \phi), \theta]D^{3}(\theta, \phi)P(\theta, \phi, \tau)dl}{|\nabla{h}(\theta, \phi)|},
    \label{eq:flux_PL}
\end{align}
where $z$ is the source redshift, $\nu (1+z)/D(\theta, \phi) = \nu'$ is the frequency in the emitting element comoving frame, ${\eta '}_{\nu'}(\nu', \theta)$ is the emitted energy per unit area, per unit frequency per unit solid angle 
and can be expressed as $\eta'_{\nu'}(\nu', \theta) = \eta'_{\nu'_0}\epsilon(\theta)S'_{\nu'}(\nu')$, where $\epsilon(\theta)$ is the structure in energy, $S'_{\nu'}(\nu')$ is the spectral shape normalized to 1 at $\nu' =\nu'_0$ and the constant $\eta'_{\nu'_0} \equiv \eta'_{\nu'}(\nu' = \nu'_0, \theta = 0)$. In Eq. \ref{eq:flux_PL} $d_L$ is the source luminosity distance\footnote{Concerning the cosmological transformations, for the sake of simplicity and since they do not impact on the final conclusions of this paper, we consider the source at $z=0$.}, $P(\theta, \phi, \tau)$ is the optical depth-dependent projection factor (with $\tau$ being the optical depth) taking into account the orientation of the emitting surface with respect to the line of sight in the optically thick regime ($P(\theta, \phi, \tau)=1$ when  the opacity is negligible), and the function $h(\theta, \phi)$ is defined as
\begin{equation}
    h(\theta, \phi) \equiv D(\theta, \phi)\Gamma(\theta)t_{\rm obs} - t_{\rm em}, 
    \label{eq:hfunction}
\end{equation}
such that:
\begin{align}
    |\nabla h(\theta, \phi)| &=  \frac{D(\theta, \phi)\Gamma(\theta)}{c}\Bigl\{\frac{1}{(\partial_\theta R)^2 + R^2}\bigl[\partial_\theta R\times\notag\\
    &\times(\cos \theta_v \cos \theta +\sin \theta_v \sin \theta)+\notag\\ &+ R(\cos \theta \cos \phi \sin \theta_v - \cos \theta_v \sin \theta)\bigr]^2+\notag\\ &+ \sin^2\theta_v \sin^2 \phi\Bigr\}^{1/2},
\end{align}

Eq. \ref{eq:dopplerfactor} and \ref{eq:flux_PL} are derived in Appendix \ref{sec:EATReq} and Appendix \ref{sec:AppendixInf}, respectively.

Throughout the paper, we will assume a \ac{SBPL} spectral shape, expressed by the function:
\begin{equation}
    S'_{\nu'}(\nu') = \frac{2}{\Bigl(\frac{\nu'}{\nu'_0}\Bigr)^{\alpha_s} + \Bigl(\frac{\nu'}{\nu'_0}\Bigr)^{\beta_s}}
    \label{eq:SBPL}
\end{equation}
where $\alpha_s$ and $\beta_s$ are the spectral indices and $\nu'_0$ is the maximum frequency of $\nu' S'_{\nu'}$ in the comoving frame, which corresponds in the observer frame to the on-axis maximum frequency $\nu_{\rm peak} \equiv D(\theta = 0, \phi = 0)\nu'_0/(1+z)$. A unique comoving--frame spectrum along the jet structure is assumed. 
 We will briefly discuss the impact of this assumptions on the spectral evolution observed during the steep-plateau phases in Sec. \ref{sec:summary}, while we remand the reader to \citet{Salafia2015} for a study of the consistency of the structured jet model with the Amati and Yonetoku correlations. 
\begin{figure*}
    \centering
    \includegraphics[width =
    \textwidth]{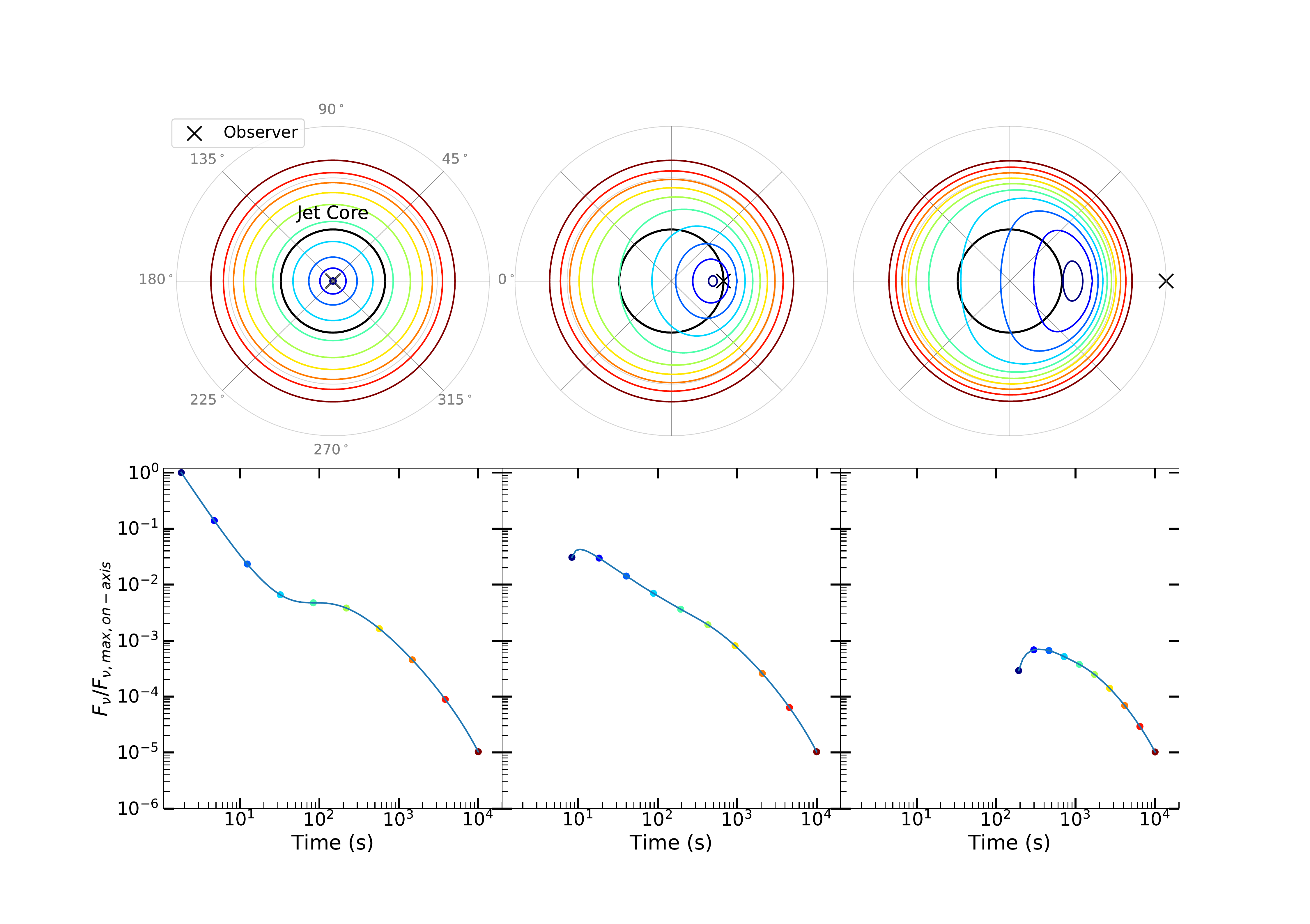}
    \caption{Equal Arrival Time Rings (top panels) and corresponding lightcurves (bottom panels) for three configurations of the viewing angle (from left to right), $\theta_v = 0^\circ$ (on-axis), $\theta_v = \theta_c$ (edge-of-the-core) and $\theta_v = 3\times \theta_c$ (off-axis).
    \emph{Top panels}: polar plot of the emitting surface. The black circle shows the physical limit of the core of the jet. The black cross shows the observer line of sight. The colored curves are the \acp{EATR} at 10 arbitrary sampling times.
    \emph{Bottom panels}: lightcurve normalized at its maximum value (corresponding to the on-axis observer at the first sampled epoch). The colored dots represent the flux emitted by the corresponding colored \acp{EATR} in the top panel. The structured jet assumed for this test case has a Gaussian strcture with $\Gamma_c = 100$, $\theta_c = \theta_E = 3^\circ$, $R_0 = 10^{15}\,\rm cm$. The comoving  emission spectrum is described by a \ac{SBPL} spectrum with $\alpha_s = 0.2$ and $\beta_s = 1.3$.}
    \label{fig:eatr_lc}
\end{figure*}

 Fig. \ref{fig:eatr_lc} shows the lightcurve (bottom panels) for three  different values of the viewing angle $\theta_{v}$ (from left to right): on-axis ($\theta_{v}=0$), edge of the jet core ($\theta_{v}=\theta_{c}$), and off axis ($\theta_{v}=3\times \theta_{c}$). For these examples we have considered a Gaussian jet with the following parameters: $\Gamma_c = 100$, $R_0 = 10^{15}\,\rm cm$, $\theta_c = \theta_E = 3^\circ$, while for the spectrum we choose a \ac{SBPL} function with a peak in the observer frame in the on-axis case at $h\nu_{\rm peak} = 100\, \rm keV$ (where $h$ is the Planck constant) and energy spectral indices $\alpha_s = 0.2$, $\beta_s = 1.3$, below and above the peak, respectively. 

The upper panels in Fig.~\ref{fig:eatr_lc} show the emitting surface in polar coordinates such that the center represents the jet axis. 
The cross represents the position of the line of sight and the black circle marks the jet core boundary $\theta_c$ (which coincides with $\theta_E$ in the particular example 
shown here). The colored curves represent the \ac{EATR} at different times. Note that for an on--axis observer (upper left panel) the \acp{EATR} are concentric circles, as expected from axisymmetry, while for off--axis observers (upper central and right panels) the \acp{EATR} become more and more distorted as $\theta_{v}$ increases, due to the breaking of axial symmetry combined with the latitudinal structure of the jet. 

The bottom rows of Fig.~\ref{fig:eatr_lc} show the lightcurves, i.e.~the flux density $F_\nu(t)$ at $h\nu = 10\,\rm keV$ normalized to the maximum value received by an on-axis observer as a function of time. Colored dots on the continuous lines correspond to the EATRs marked with the same color coding in the corresponding top panel. 

In the case of an on--axis observer, shown in the left top and bottom panels of Fig.~\ref{fig:eatr_lc}, the initial (from  0 to $\sim$50 s) steep decay of the lightcurve corresponds to emission produced within the core of the jet (as predicted by the analytical solution of  \citealt{Kumar2000}).
The following plateau phase (from $\sim$50 to $\sim$300 s) is instead due to photons emitted by regions just outside of the jet core: these photons are seen by the on-axis observer due to the increasing opening angles of the beaming cones in these regions characterized by a decreasing $\Gamma$.

The transition between the steep decay and the plateau phase becomes less evident as the line of sight moves to larger off-axis angles (central and right column plots of Fig.~\ref{fig:eatr_lc}). 
For $\theta_v = \theta_c$ (central column), after the peak a transition from a steep to a shallower decay can still be seen, although much less evident than in the on-axis configuration. From the central and right top panels of Fig.~\ref{fig:eatr_lc} one can see that, similarly to the on-axis case, the steep decay is produced by the emission from \acp{EATR} lying (at least partially) inside the core of the jet. The the plateau is produced by \acp{EATR} just outside the core of the jet. Fig.~\ref{fig:eatr_lc} shows that by increasing the viewing angle (i.e.~more off-axis lines of sight) the distinction between the steep decay and the plateau sections of the lightcurve becomes progressively less marked.

Another feature of the off-axis lightcurves is the fainter peak flux density (e.g.~two and four orders of magnitudes for $\theta_v = \theta_c$ and $\theta_v = 3\theta_c$, respectively) with respect to that measured by the on-axis observer.
This effect is due to the fact that, for $\theta_v = \theta_c$ and $\theta_v = 3\theta_c$, the core of the jet, where $\epsilon(\theta)$ and $\Gamma(\theta)$ reach their maximum values, is beamed outside of the line of sight.

Finally, the time of arrival of the first photon is delayed for increasing viewing angle, such that for $\theta_v = 0^\circ, \theta_c, 3\theta_c$ it arrives respectively at $\sim 2\, \rm s, 10 \, \rm s$ and $200\, \rm s$ after the jet launch time. 

\subsection{Finite Duration Emission}
\label{sec:finite}

In the previous section, similarly to O20, we considered the simplifying assumption of an impulsive emission episode in computing its \ac{HLE} lightcurve.
In this section we generalize our calculation to the case of a finite time interval during which the energy is released. 
To this aim we approximate the emission episode with a temporal profile $I_\nu(t)$ which is the sum of $N$ infinitesimal pulses. Each infinitesimal pulse is treated as in the impulsive approximation. The $N$ pulses are then summed in such a way that the total emitted energy is conserved (a more detailed description of the method is provided in Appendix \ref{sec:AppendixFinite}).

Fig.~\ref{fig:finite} shows the lightcurve  obtained with the finite-duration emission case for an on-axis observer. Also for this example we consider a Gaussian jet with a core half-opening angle $\theta_c = \theta_E = 3^\circ$ and $\Gamma_c = 100$. The emission starts at a radius $R_0 = 10^{13}\,\rm cm$ and its temporal profile is described by a step function sampled $N = 105$ times and lasting $\Delta T = 10^{5} \rm s$ in the \ac{CoE} frame, which 
corresponds to $\Delta T_{\rm obs} \simeq \Delta T/(2\Gamma^2_c) = 5\, s $ in the observer frame.

In Fig.~\ref{fig:finite} the black solid line represents the total flux density, while the dashed colored lines show the contributions to the emission of the $N$ discrete pulses, which henceforth we refer to as `partial lightcurves'\footnote{For a better visualization, we show only some selected $n = 13$ out of $N = 105$ partial lightcurves, shifting in the figure their fluxes by a factor $N/n$ to properly represent the contribution of the emission at different radii.}. 
The emitting surfaces of all pulses have the same structure but they correspond to different emission radii. We labelled each surface with the parameter $\Delta R_0 \equiv R_i(\theta = 0) - R_0$, where $R_i (\theta)$ is the radius of jet core for the i-th surface. Each emitting surface is characterized by the constant ($\eta'_{\nu'_0,i}$ in Eq.~\ref{eq:norm_finite}) which 
we assume - in this particular example - constant with time. 
We assume for simplicity that the spectrum and the energy profile $\epsilon(\theta)$ do not change within the pulse. 

We can see from Fig. \ref{fig:finite} that also in the more realistic case of a finite duration emission episode the lightcurve morphology is preserved in the on--axis case, because the total lightcurve is dominated by the very late partial lightcurves. The presence of the steep decay-plateau structure in the on-axis case found by O20 is confirmed when a more realistic finite duration emission is considered.  We have verified that also in the off-axis case the final partial lightcurves shape the total emission for most of the time, with the exception (like in the on-axis case) of  first phase, when the contribution of the early time emission is non-negligible. 

It is quite important to  notice  -- as shown in Fig. \ref{fig:finite} --  that from our model the duration of the plateau is not necessarily determined by the radius  of  the  prompt  emission.  The  only requirement  to  have  a  long  lasting  plateau  is  that the portion of the jet outside of the core, and not necessarily the entire surface,  is  emitting  at  large  radii.  We  can  have a  scenario  in  which  the  prompt  emission  is  generated  at small radii (\emph{e.g.} $10^{13}\, \rm cm$) by the core of the jet, but while the core switches off, the portion of the surface outside the core is still emitting (or  it  switches  on  and  off with  a  delay  with  respect  of  the core) when the outflow  has  reached  a  larger  distance  (\emph{e.g.} $10^{15}\,\rm cm$). This scenario could in principle accommodate a rapid variability of the prompt emission together with a long lasting plateau  without  invoking mini-jets \citep{KumarPiran2000}.

\begin{figure*}
    \centering
    \includegraphics[width=\textwidth]{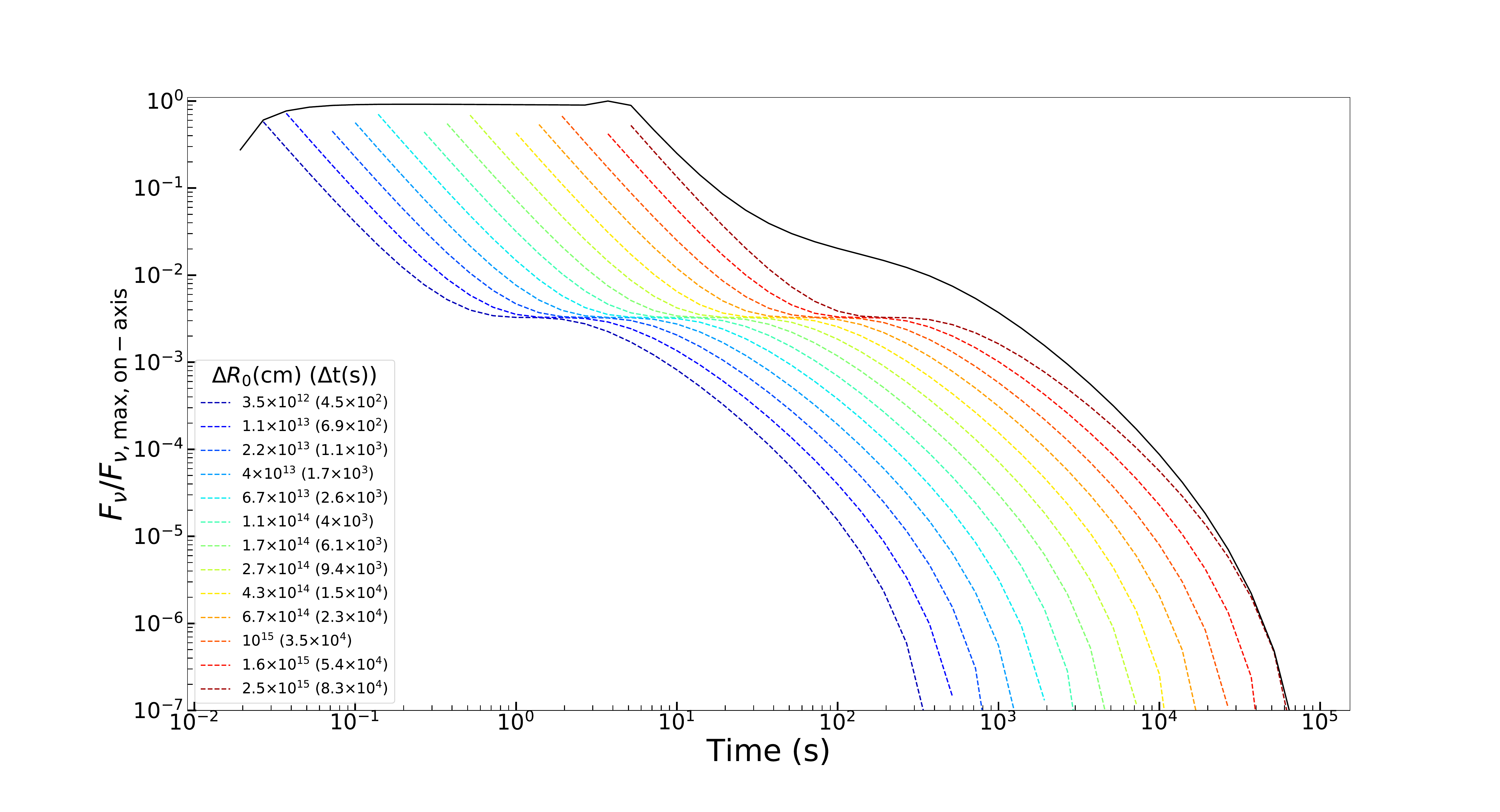}
    \caption{Lightcurve for finite duration of the emission. The case refers to an on-axis observer. The dashed colored  lightcurves corresponding to $n = 13$ selected infinitesimal pulses out of the total $N = 105$ composing the finite duration emission episode. Partial fluxes have been multiplied by a factor $N/n$, in order to visualize better the contribution to the final emission at different radii. Colors identify the $\Delta R_{0,i}$ ($\Delta t$) with respect to the initial $R_0$ ($t_{\rm em}$) at which the first pulse is emitted. The black solid line is the final lightcurve, made by the sum of the partial ones. We assume a Gaussian jet with parameters $\Gamma_c = 100$, $\theta_c = \theta_E = 3^\circ$, $R_0 = 10^{13}\,\rm cm$ and the pulse duration in the \ac{CoE} frame $\Delta T = 10^5\,\rm s$. A \ac{SBPL} spectrum is with $\alpha_s = 0.2$ and $\beta_s = 1.3$ is assumed. }
    \label{fig:finite}
\end{figure*}

\section{Opacity}
\label{sec:opacity}

O20 assumed that the entire emission region is transparent. This assumption is allowed when the emission radius $R_0$ is large. Since a large radius is also required to have a long lasting plateau, this justifies O20 approximation. 
Here we consider a more realistic scenario, where both the effect of the latitude-dependent opacity, its variation with the radius, and the finite duration emission are included.

It is well known that ultra--relativistic motion is required in \acp{GRB} to solve the so called compactness problem \citep{Ruderman1975,Schmidt1978}. Originally, this argument is applied considering the main source of opacity for high energy photons, i.e. pair production \citep{Piran1999,Lithwick2001}. Recently, this argument was also applied to constrain the viewing angle \citep{Matsumoto2019}. Considering that our study aims at interpreting the soft X-ray emission (0.1-10) keV, we consider hereafter only the Thomson scattering as the main source of opacity.

In Appendix~\ref{sec:AppendixOpacity}, we calculate the expression for the $\theta$-dependent optical depth, which reads

\begin{align}
    \tau(\theta) = &\frac{Y_e \sigma_T L_{\rm K, ISO}(\theta)}{4\pi m_p \beta(\theta) [1 + \beta(\theta)]c^3\Gamma^2(\theta)[\Gamma(\theta) -1]}\notag\\&\times\,\Bigl\{\frac{1}{R(\theta)} - \frac{1}{[1+\beta(\theta)]\Gamma^2(\theta)\Delta R(\theta) + R(\theta)}\Bigr\},
    \label{eq:optical_depth}
\end{align}
where $Y_e$ is the electron fraction, $\sigma_T = 6.65\times 10^{-25}\, \rm cm^2$ the Thomson scattering cross- section, $m_p = 1.67\times10^{-24}\, \rm g$ the mass of the proton, $L_{K, ISO}(\theta)$ the kinetic isotropic equivalent luminosity, $\Delta R(\theta)=\beta(\theta)c\Delta T_\mathrm{engine}$ is the width of the emitting region, and $\Delta T_{\rm engine}$ is the duration of the central engine. The kinetic luminosity profile is set by the relation $dE_K/d\Omega \propto \beta^2(\theta)\epsilon(\theta)$ (O20), such that we can write
\begin{equation}
    L_{\rm K, ISO}(\theta) = \frac{E_{\rm K, ISO}}{\Delta T_{\rm engine}}\beta^2(\theta)\epsilon(\theta),
\end{equation}
where $E_{\rm K, ISO}$ is the isotropic equivalent kinetic energy. For these two parameters we take the reference values $E_{\rm K, ISO} = 10^{54}\, \rm erg$ and $\Delta T_{\rm engine} = 30\, \rm s$, appropriate for long \acp{GRB}.

\begin{figure*}
    \centering
    \includegraphics[width=\textwidth]{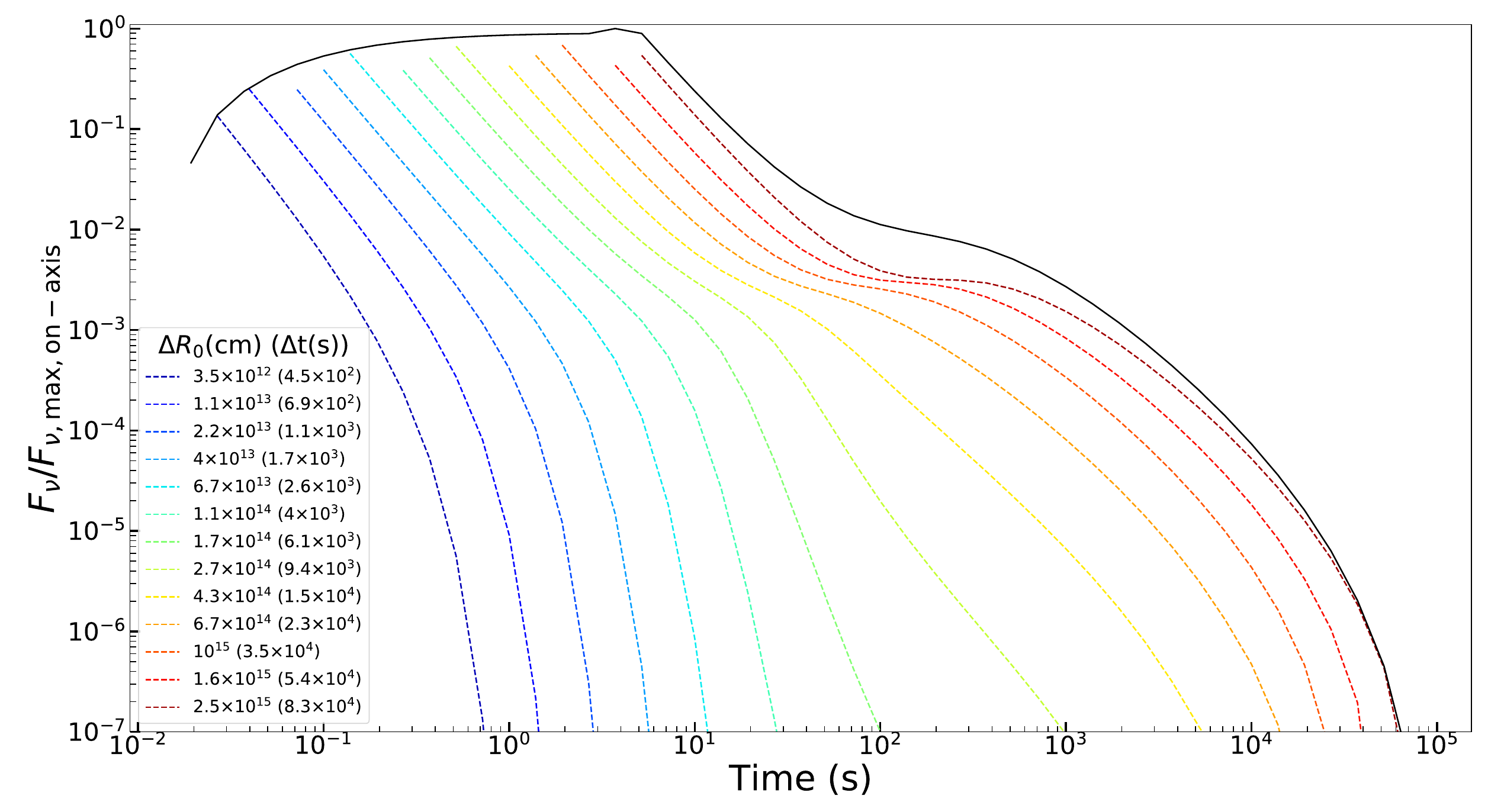}
    \caption{Same as Figure \ref{fig:finite} with the effect of a non-negligible and latitude-dependent opacity. The parameters for the optical depth are: $Y_e = 1$, $E_{\rm K, ISO} = 10^{54}\,\rm erg$ and $\Delta T_{\rm engine} = 30\,\rm s$.}
    \label{fig:finite_with_opacity}
\end{figure*}

The optical depth of Eq.~\ref{eq:optical_depth} multiplies the integrand in the RHS of Eq.~\ref{eq:flux_PL} (\ref{eq:flux_Band}) through the multiplicative factor $f(\theta) = \exp[-\tau(\theta)]$, which suppresses the emission in the regions where $\tau(\theta) > 1$.
The absorption effect is shown in Fig. \ref{fig:finite_with_opacity}. 
The partial lightcurves emitted at small radii (dark blue dashed lines) are strongly absorbed due to the large opacity along most of the emission surface and, particularly, outside the jet core where $\Gamma(\theta)$ decreases with the increasing angle from the jet axis. However,
the contributions to the emission produced at larger radii suffer  a  smaller  absorption  (as  shown by the orange-red  lines).  Overall,  also  considering  the  effect  of  the angular dependence of the opacity,  the appearance of the  plateau for the  structured jet parameters considered in this example is preserved.

In the following we will restrict ourselves to the infinitesimal emission case for which we consider the damping due to the opacity. This choice is motivated by the evidence that the steep decay and plateau phases mostly depend on the very late ``partial'' lightcurves (see Fig. \ref{fig:finite}), which are ascribed to the final phases of the emission of a finite pulse.  
However, if the emission ceases at sufficiently small radii (\emph{e.g.} $R_0 \sim 10^{14}\,\rm cm$) the opacity plays a dominant role in shaping the lightcurve (see Fig. \ref{fig:finite_with_opacity}) and its effect needs to be considered. 

It is also worth noticing that when the opacity is non-negligible, the rest-frame spectrum in the regions outside of the core is unlikely to be exactly equal to the core spectrum (due to e.g. absorption and Comptonization of photons).  Despite this, for simplicity, we consider the case of a uniform unique comoving frame spectrum along the jet surface, since at large radii opacity-driven spectral corrections are expected to be negligible.

\section{Results}
\label{sec:results}
\subsection{Parametric Study}

Here we investigate the lightcurve dependence on model parameters. We define a fiducial model by choosing a particular set of parameter values, and then vary one parameter at a time. More specifically, our fiducial model is the same adopted in Fig.~\ref{fig:eatr_lc} -- namely $\Gamma_c = 100$, $R_0 =10^{15}\,\rm cm$, $\theta_c = \theta_E = 3^\circ$ and a Gaussian structure -- with a \ac{SBPL} function spectrum with indices $\alpha_s = 0.2$ and $\beta_s = 1.3$.  Instead, the lightcurves arising from a power-law jet structure are discussed in Appendix \ref{sec:Appendix-PL-structure}. 

Fig.~\ref{fig:case_a} shows the lightcurves of our fiducial model (solid lines) for three values of the viewing angle $\theta_{\rm v}=[0, 1, 3]\times\theta_c$, which correspond to the on-axis, edge and off-axis view and are represented in blue, orange and green, respectively. 
In the same figure we report also two further models characterized by the same $R_0$ and the same spectrum: the dashed curves are obtained by increasing the core half-opening angle (i.e. $\theta_c = \theta_E = 5^\circ$), while the dotted curves are obtained by increasing the core Lorentz factor (i.e. $\Gamma_c = 200$).  
  All the lightcurves are expressed as $F_\nu(t_{\rm obs})$ at a reference (observer frame) energy of $h\nu = 10\, \rm keV$ and each curve is normalized to the maximum of the on-axis lightcurve. 
Differently from the previous figures, here we refer the time axis to the time of arrival of the first photon, which, for a better visualization, we set at $10^{-1}$ s.
This choice is motivated by the likely possibility that, in the upcoming years, a large fraction of off-axis \acp{GRB} will be directly detected by high-energy surveys and we will lack a reference initial time that could be provided only by a \ac{GW} or neutrino trigger.

\begin{figure*}
    \centering
    \includegraphics[width = \textwidth]{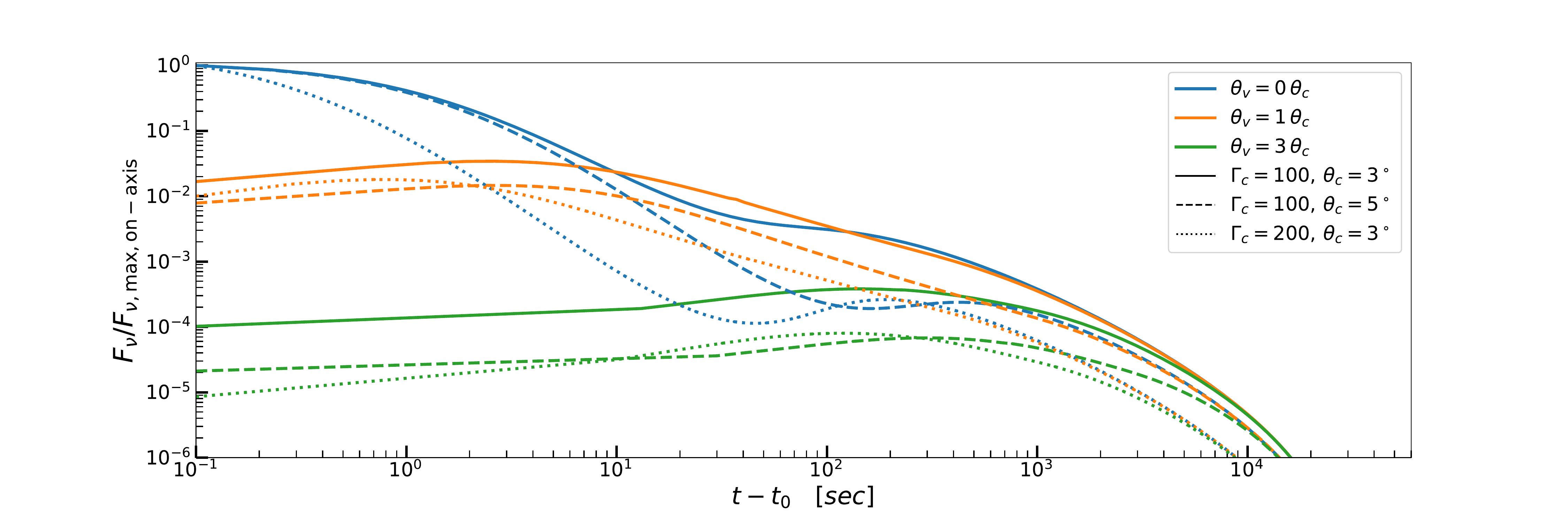}
    \caption{Lightcurves obtained varying $\Gamma_c$ and $\theta_c$ at fixed $R_0 = 10^{15}\,\rm cm$, at different viewing angles. Lightcurves corresponding to viewing angles $\theta_v = 0,\, \theta_c,\, 3\theta_c$ are shown in blue, orange and green, respectively. Continuous curves denote the fiducial model, with $\Gamma_c = 100$, $\theta_c = 3^\circ$. The structure is Gaussian and the spectrum is a \ac{SBPL} function with $\alpha_s = 0.2$ and $\beta_s = 1.3$. The lightcurves are those at $\nu = 10\, \rm keV$, normalized to the maximum value of the on-axis case and the time is relative to the time of arrival of the first photon. For a better visualization, we translated the time such that the first photons arrive at $10^{-1}\, s$.}
    \label{fig:case_a}
\end{figure*}

\begin{figure*}
    \centering
    \includegraphics[width = \textwidth]{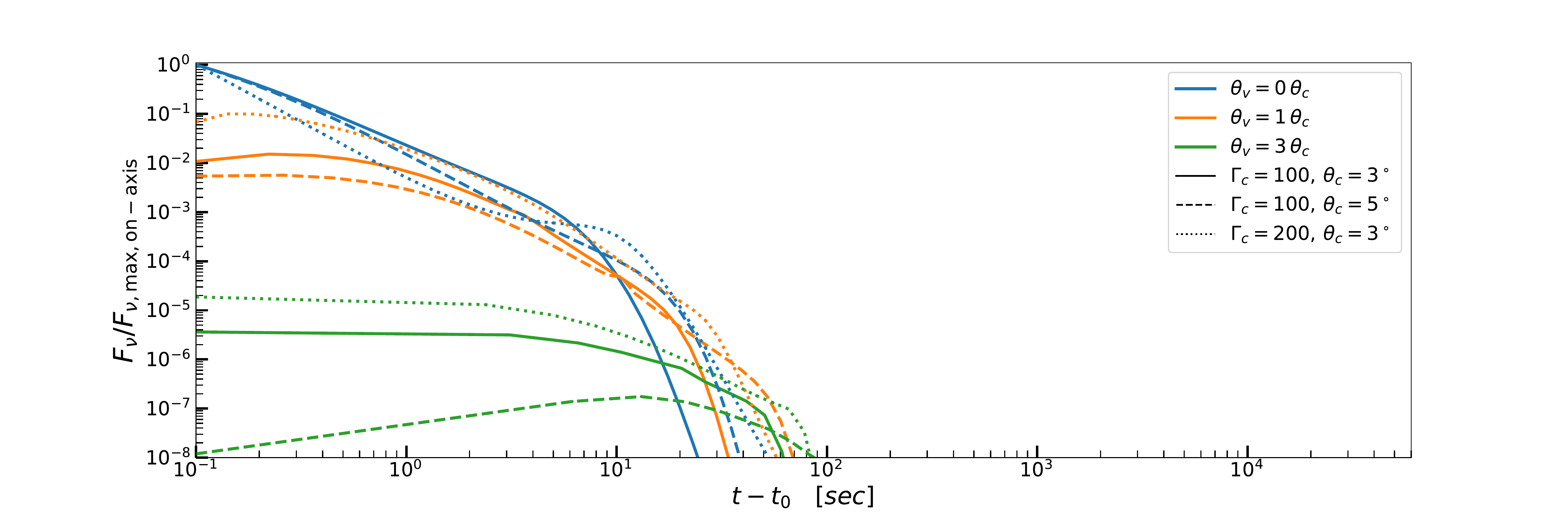}
    \caption{Same as Fig.~\ref{fig:case_a} but with $R_0 = 10^{14}\,\rm cm$. In the off-axis case ($\theta_v = 3\theta_c$) the emission is quenched as described in the text (note the different y-axis scale with respect to Fig. \ref{fig:case_a}).}
    \label{fig:case_b}
\end{figure*}

As shown by the dotted curves in Fig.~\ref{fig:case_a}, the Lorentz factor of the core mainly affects the on-axis curve. A higher Lorentz factor corresponds to a steeper initial decay, and produces a bump or rebrightening at later times as the emission from outside of the core becomes dominant. Moreover, a larger $\Gamma_c$ causes the emission to be fainter than the equivalent model with lower $\Gamma_c$, and in general the whole lightcurve fades more rapidly. This behaviour can be explained by the narrower relativistic beaming of the radiation, that suffers, consequently, a faster angular dependent de-beaming effect. 
When the \acp{EATR} cross the core boundary, the increase of the flux due to the Doppler factor overcomes the de-beaming due to the increasing latitude, producing a rebrightening instead of a plateau. The off-axis lightcurves instead show less prominent differences with respect to the equivalent ones in the fiducial model. The lightcurve corresponding to the edge-of-the-core viewing angle does not show a plateau or rebrightening.
Interestingly, since the edge-of-the-core lightcurve shows a constant decay while the on-axis lightcurve has a rebrightening, under the assumption that the lightcurves change smoothly with $\theta_v$, we may conclude that there must exists a set of $\theta_v$ values, included in the interval $(0, \theta_c)$, for which the lightcurves present a plateau. This means that, at least in this case, observers with $\theta_v < \theta_c$ but not perfectly aligned with the axis will see a plateau in the lightcurve.

Instead, an on-axis model with a wider core opening angle (dashed line of Fig.~\ref{fig:case_a}) is characterized
by a longer lasting steep decay and a longer lasting and fainter plateau with respect to the fiducial model. When this is accompanied with a larger $R_0$ the plateau can last as long as few $10^4\,\rm s$, as shown in O20.
 The off-axis lightcurves in this case are qualitatively similar to those of the other two models.

In Fig. \ref{fig:case_b} we consider the same models as in Fig. \ref{fig:case_a} but assuming a smaller emitting radius of $R_0 = 10^{14}\,\rm cm$. As in Fig. \ref{fig:case_a}, the blue, orange and green colors denote lightcurves at $\theta_v =[0, 1, 3]\times \theta_c$ respectively, while the continuous, dashed and dotted lines identify different combinations of $\Gamma_c$ and $\theta_c$. The shape of the lightcurves changes substantially: the steep decay is present in all the cases while the plateau disappears. The only exception with a short plateau phase is with $\Gamma_c = 200$ (dotted line). By comparing this result with Fig. \ref{fig:finite_with_opacity} we can understand that this effect is due to the opacity which suppresses completely the emission of the jet wings. 
If the opacity effect were ignored, the on-axis lightcurve would retain the steep decay-plateau structure, with a shorter duration plateau compared to the fiducial model (see for example Fig. \ref{fig:finite}). It is worth noticing that the opacity-driven suppression of the plateau can be used to constrain the parameters entering in Eq. \ref{eq:optical_depth}. For example, if we are confident on our knowledge of the parameters describing the jet structure (e.g $\Gamma(\theta)$) and those describing  , the microphysics (e.g. $Y_e$), the presence of a plateau and its duration can determine the radius of emission $R_0$. Off-axis lightcurves ($\theta_{v}=3\theta_{c}$ - green line) are strongly suppressed (note the different scale of the vertical axis in Fig. \ref{fig:case_a} and Fig. \ref{fig:case_b}) because the radiation beamed towards the observer is quenched by the large opacity and the radiation from the core (less absorbed), which would suffer less absorption, is strongly de-beamed.

\subsection{Comparison with the off-axis forward shock emission}

O20 modeled the X-ray and optical lightcurves of GRB 100906A as the sum of an on-axis \ac{HLE} and afterglow (forward-shock) emission. In this work, we calculate the off-axis lightcurve of that model, in order to investigate how the \ac{HLE} and forward-shock emission compare when the source is observed off-axis. In particular, we are interested in addressing whether one of the two components dominates over the other, or the two emissions remain well separated in time. We assume the same parameters of the model used in O20 to describe the jet; a Gaussian structure, $\Gamma_c = 160$, $R_0 = 3\times 10^{15}\, \rm cm$ and $\theta_c = 3.3^\circ$ and $\theta_E = 4.4^\circ$. For the afterglow model we consider a forward shock with ambient density $n = 15\,\rm cm^{-3}$, shock-accelerated electron power-law index $p = 2.1$, post-shock internal energy density fraction shared by the accelerated electrons $\epsilon_e = 0.03$, fraction shared by the magnetic field $\epsilon_B = 2 \times 10^{-2}$  and finally $E_{\rm K, ISO} = 5\times10^{53}\,\rm erg$. The off-axis forward shock model is described in \citet{Salafia2019}. The spectrum of the \ac{HLE} here slightly differs from the one used in O20. While O20 used a doubly smoothly broken power-law spectrum, here we use a simpler \ac{SBPL}, with $\alpha_s = 0.67$,  $\beta_s = 3.0$ and $h\nu_{\rm peak}= 100\,\rm keV$.

\begin{figure}
    \centering
    \includegraphics[width = \columnwidth]{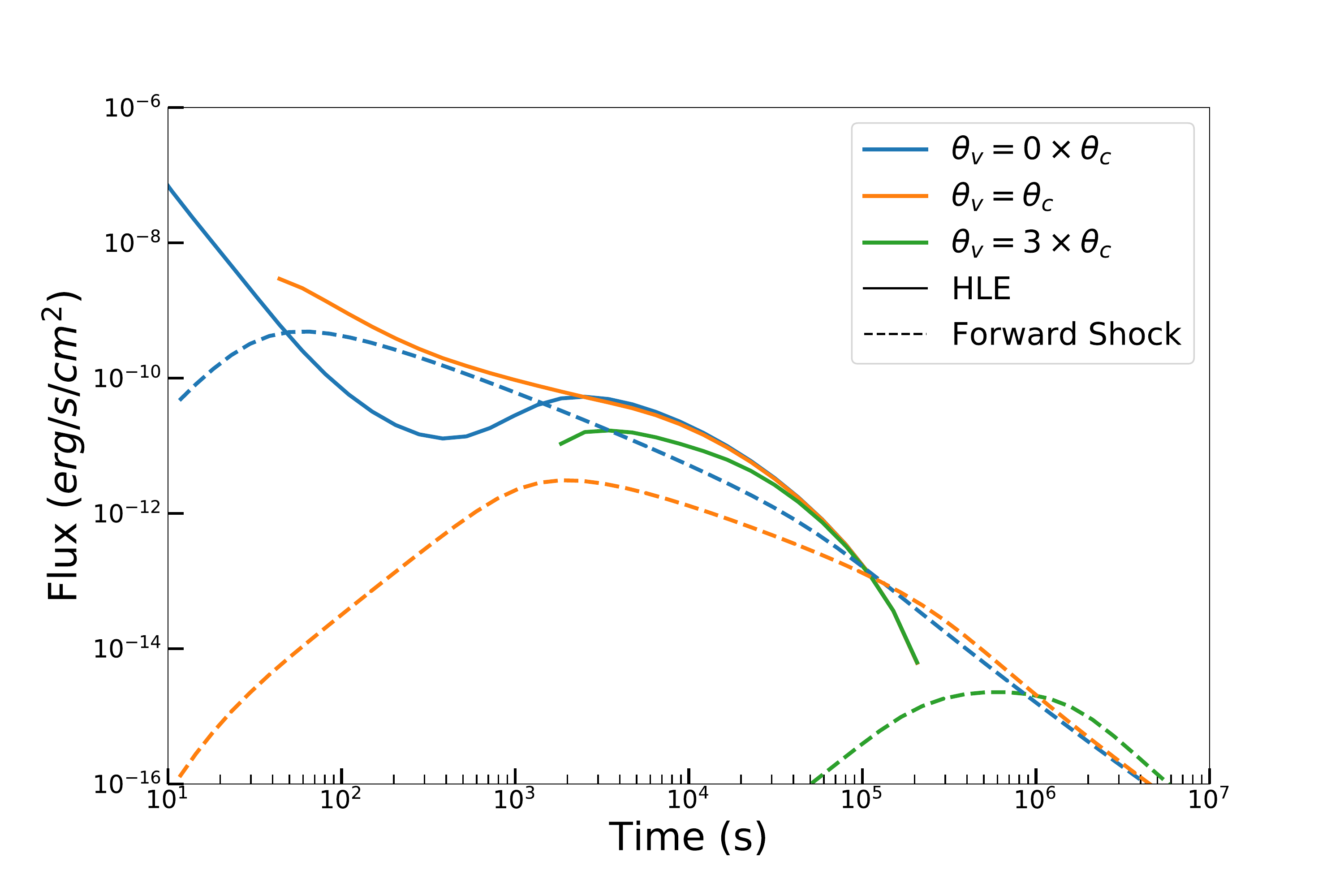}
    \caption{The on-axis ($\theta_v = 0\times \theta_c$), edge ($\theta_v = \theta_c$) off-axis view ($\theta_v = 3\times\theta_c$) of \ac{HLE} (continous line) and forward shock (dashed lines) models used in O20 to fit the lightcurve of GRB 100906A. The lightcurves are integrated in the energy range $0.5-10\,\rm keV$. \ac{HLE} model parameters: $\Gamma_c = 160$, $R_0 = 3\times 10^{15}\, \rm cm$ and $\theta_c = 3.3^\circ$ and $\theta_E = 4.4^\circ$, spectrum: \ac{SBPL}, $\alpha_s = 0.67$,  $\beta_s = 3.0$ and $\nu_{\rm peak}= 100\,\rm keV$, jet structure: Gaussian. Forward shock model parameters: $n = 15\,\rm cm^{-3}$, $p = 2.1$, $\epsilon_e = 0.03$, $\epsilon_B = 2 \times 10^{-2}$, $E_{\rm K, ISO} = 5 \times 10^{53}\,\rm erg$.}
    \label{fig:FScomparison}
\end{figure}

Fig. \ref{fig:FScomparison} shows the corresponding \ac{HLE} and the forward-shock lightcurves with continuous and dashed lines, respectively. In blue, orange and green colors are represented the on-axis ($\theta_v = 0\times \theta_c$), edge-of-the-core ($\theta_v = \theta_c$) and off-axis views ($\theta_v = 3\times\theta_c$), respectively.
We can see that the off-axis forward shock emission does not outshine the \ac{HLE}, because it peaks when the latter has already faded. Moreover, the forward shock emission is more suppressed than the \ac{HLE} for increasing viewing angles. This means that the off-axis \ac{HLE} emission may be observed even when the forward shock cannot. 
The chosen example, GRB100906A,  is  a  long \ac{GRB}, but our results hold also for short \acp{GRB}.
While short \acp{GRB} are characterized by a factor $\sim10$ fainter afterglows with respect to long \acp{GRB} \citep{Berger2014}, a smaller factor is expected for the \ac{HLE} corresponding to the difference in prompt emission luminosity of long and short \acp{GRB} \citep{DAvanzo2014}.

This has important implications in the multi-messenger context; for example, in the case of \ac{EM} follow-up of \ac{GW} from a \ac{BNS} merger the \ac{HLE} emission of the associated short \ac{GRB} can be detected off-axis, with X-ray observations made within a few hours from the \ac{GW} trigger. 
Furthermore, wide field X-ray telescopes, such as eROSITA and Einstein Probe, and mission concept such as the THESEUS-\ac{SXI} could detect the off-axis \ac{HLE} of long and short \acp{GRB}, irrespective of a \ac{GW} trigger.

\begin{figure}
    \centering
    \includegraphics[width = \columnwidth]{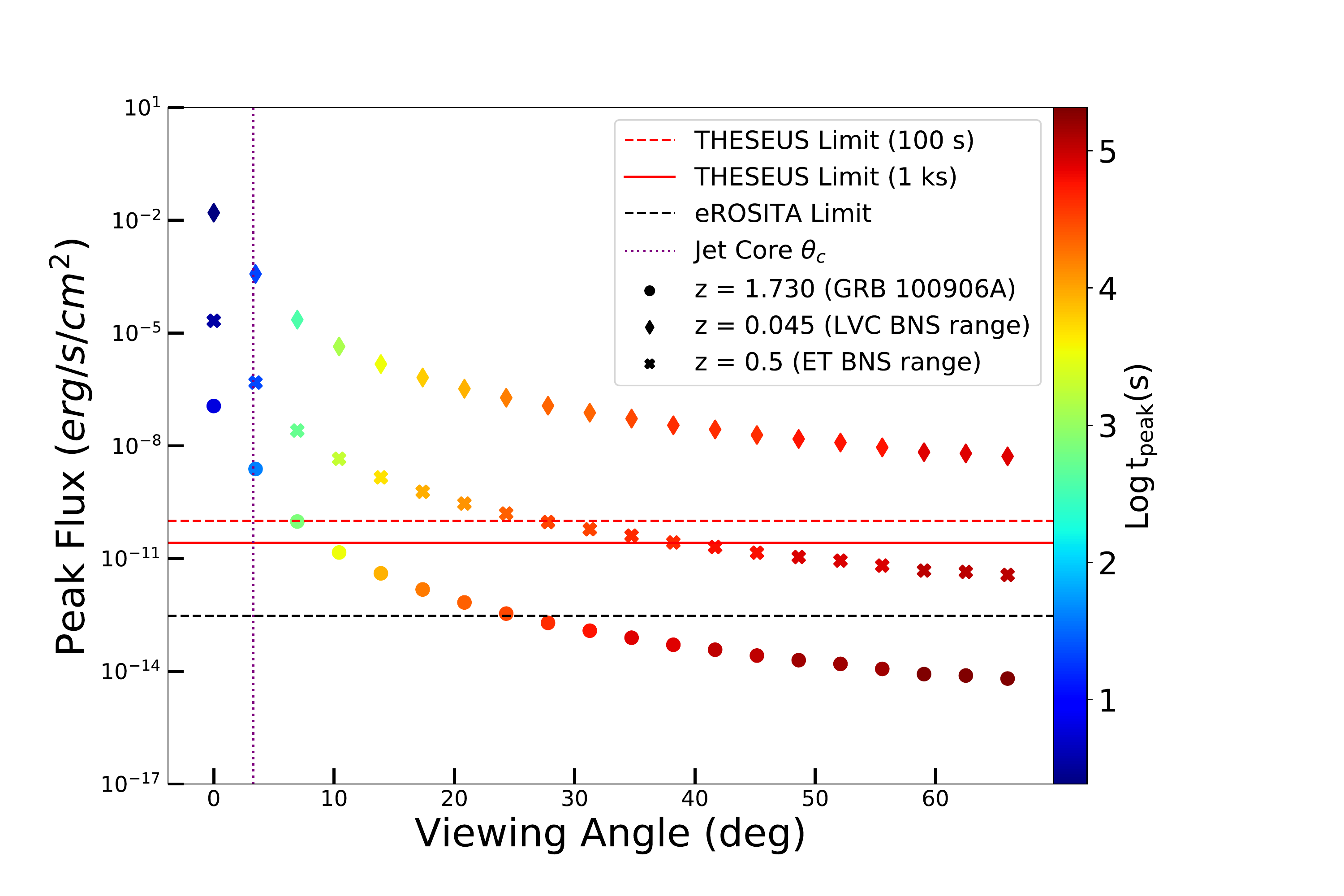}
    
    \caption{Peak flux vs viewing angle of the \ac{HLE} of GRB 100906A assuming the same structure derived in O20. Circles, diamonds and crosses represent the same model at three different redshifts: z = 1.730 (actual redshift of GRB 100906A), z = 0.045 (Advanced LIGO and Virgo range for \ac{BNS} mergers) and z = 0.5 (\ac{ET} range for \ac{BNS} mergers), respectively. The color code denotes the logarithm of the time of the peak in the observer frame. Red continuous and red dashed lines mark the flux limit of THESEUS-\ac{SXI} integrated in $10^3\,\rm s$ and $10^2\,\rm s$, respectively. Black dashed line marks the flux limit of eROSITA integrated in $40\,\rm s$. Verical dotted line marks $\theta_c$. }
    
    \label{fig:FluxAngle}
\end{figure}

In order to illustrate the capability of these instruments to detect the off-axis \acp{HLE}, Fig. \ref{fig:FluxAngle} shows the peak flux of our model for GRB100906A at different viewing angles by placing the source at different redshifts: $z = 1.730$, the actual redshift of this event, $z = 0.045$, corresponding to the luminosity distance of $200\,\rm Mpc$ that we consider as a fiducial limit for the detection of a \ac{BNS} by the \ac{LVC} \citep{Abbott2018distance}, and $z = 0.5$, the expected BNS range for the \ac{ET} \citep{ETdesignstudy, Maggiore2020}. The \ac{ET} range corresponds to the redshift up to which the $\,60\%$ of the sources are expected to be detected by the interferometer \citep{Maggiore2020}. These three $z$ are represented by circles, diamonds and crosses, respectivelly. The color code of the symbols, denotes the time at which the peak occurs in the observer frame. The red solid and dashed lines, show the limiting fluxes of $2.6\times 10^{-11}\, \rm erg/s/cm^2$ ($10^3\,\rm s$ of exposure time) and $10^{-10}\, \rm erg/s/cm^2$ ($10^2\,\rm s$ of exposure time) for THESEUS-SXI\footnote{A source is expected to remain in the THESEUS-SXI field of view for about $100-1000\,\rm s$, considering the instrument in survey mode.} \citep{Theseus2018}, respectively, while the black dashed line shows the limiting flux of $3\times 10^{-13}\, \rm erg/s/cm^2$ ($40\, \rm s$ of exposure time) for eROSITA \citep{KaSaSu2012}. Finally, the vertical purple dotted line marks $\theta_c$. We can see that
both THESEUS and eROSITA are able to observe the \ac{HLE} at the lowest $z$ even for very large viewing angles. However, increasing the redshift reduces the viewing angle up to which an observer can detect the \ac{HLE}; at $z\sim$1.730, THESEUS ($100\,\rm s$ exposure time) and eROSITA would detect this event only if the viewing angle is smaller than $\sim$7$^{\circ}$ or $\sim 23^{\circ}$, respectively. Nevertheless, the larger field of THESEUS-\ac{SXI} could compensate the lower sensitivity with respect to  eROSITA.  

Finally, we mention also that the modelling of the forward shock is important in constraining the jet structure with the aim of robustly estimating the contribution of the \ac{HLE}. In particular, \citealt{Beniamini2020} showed that, rather independently from parameters $\epsilon_B$, $\epsilon_e$ and $n$, the jet structure can be constrained from the afterglow lightcurves. 

\subsection{The Case of GW170817}

The first \ac{EM} signal associated with GW170817  was the faint short GRB 170817A detecteded by FERMI-GBM \citep{Goldstein2017} and INTEGRAL \citep{Savchenko2017}  $1.76\,s$ after the merger. The GRB had  a peak isotropic equivalent luminosity of $1.6\times10^{47}\, \rm erg/s$ \citep{Abbott2017b}.
No counterpart was detected by the early X--ray follow up campaign. MAXI, the first to observe in the X-ray band at $4.8\,\rm hr$, provided an upper limit of $8.6\times 10^{-9}\,\rm erg/s/cm^2$ in the energy range of ($2-10\,\rm keV$). Subsequent observations by MAXI (at $6.2,\, 12.0,\, 13.7\,\rm hr$) provided other limits to the X--ray flux ($7.7\times10^{-8}, \, 4.2\times 10^{-9}, 2.2\times 10^{-9}\,\rm erg/s/cm^2$) \citep{Sugita21555, Abbott2017a}. 
\emph{Swift} observed the position of the optical counterpart AT2017gfo \citep{Coulter2017} at $14.9\,\rm hr$ after the merger obtaining an upper limit to the X--ray flux of 
$2.74\times10^{-13}\,\rm erg/s/cm^2$ ($0.3-10)\,\rm keV$ \citep{Evans2017, Abbott2017a}. 

Early Chandra observations at $2.2\,\rm days$ after the merger \emph{Chandra} did not detect any source \citep{Troja2017Nature}. Eventually, 8.9 days after the merger Chandra discovered the X-ray counterpart with an estimated flux in the range $(0.3-10)\,\rm keV$ of $(4 \pm 1.1)\times 10^{-15}\,\rm erg/s/cm^2$ \citep{Troja2017Nature}. From this date the source emission, as observed in the X-ray, optical and radio band, slowly increased with time ($\propto t^{0.8}$, \citealt{MooleyNakar2018}) until it peaked at $\sim 150\, \rm days$ after the merger \citep{DobieKaplan2018, Davanzo2018, Margutti2018}. Finally, the lightcurve showed a rapid declining phase \citep{DobieKaplan2018, AlexanderMargutti2018}. 
A structured jet \citep{Lamb2019,Margutti2018,Lazzati2018,Gottlieb2018} is able to interpret the multi-wavelength lightcurve and the proper motion \citep{Mooley2018} and size constraints \citep{Ghirlanda2018} observed in the radio band.

Here we verify that our off-axis \ac{HLE} model is consistent with the upper limits from the early X-ray observations of GW170817. To this aim, we assume the jet parameters derived from the combined modelling of the afterglow emission, source proper motion and size constraints by \citet{Ghirlanda2018}. The jet is assumed to be observed off-axis at a viewing angle $\theta_v = 15^\circ$ and to have a power-law structure with core parameters  $\Gamma_c = 251$, $\theta_c = \theta_E = 3.4^\circ$, $E_{\rm K, ISO} = 2.5\times 10^{52}\, \rm erg$. The angular profiles of the bulk Lorentz factor and of the energy are modelled with different power-law slopes, i.e. $k_c = 3.5$ and $k_E = 5.5$ respectively. 
We consider two possible values for the emission radius $R_0$, corresponding to $R_0 = 10^{13}\, \rm cm$ and $R_0 = 2\times10^{15}\,\rm cm$ and we model the \ac{HLE} in these two limiting cases. We consider a \ac{SBPL} function with $h\nu_{\rm peak} = 1500\,\rm keV$ ($h\nu'_0 = 3\,\rm keV$ in comoving frame), $\alpha_s = 0.5$ and $\beta_s = 1.3$ as a reference spectrum. 

\begin{figure}
    \centering
    \includegraphics[width = \columnwidth]{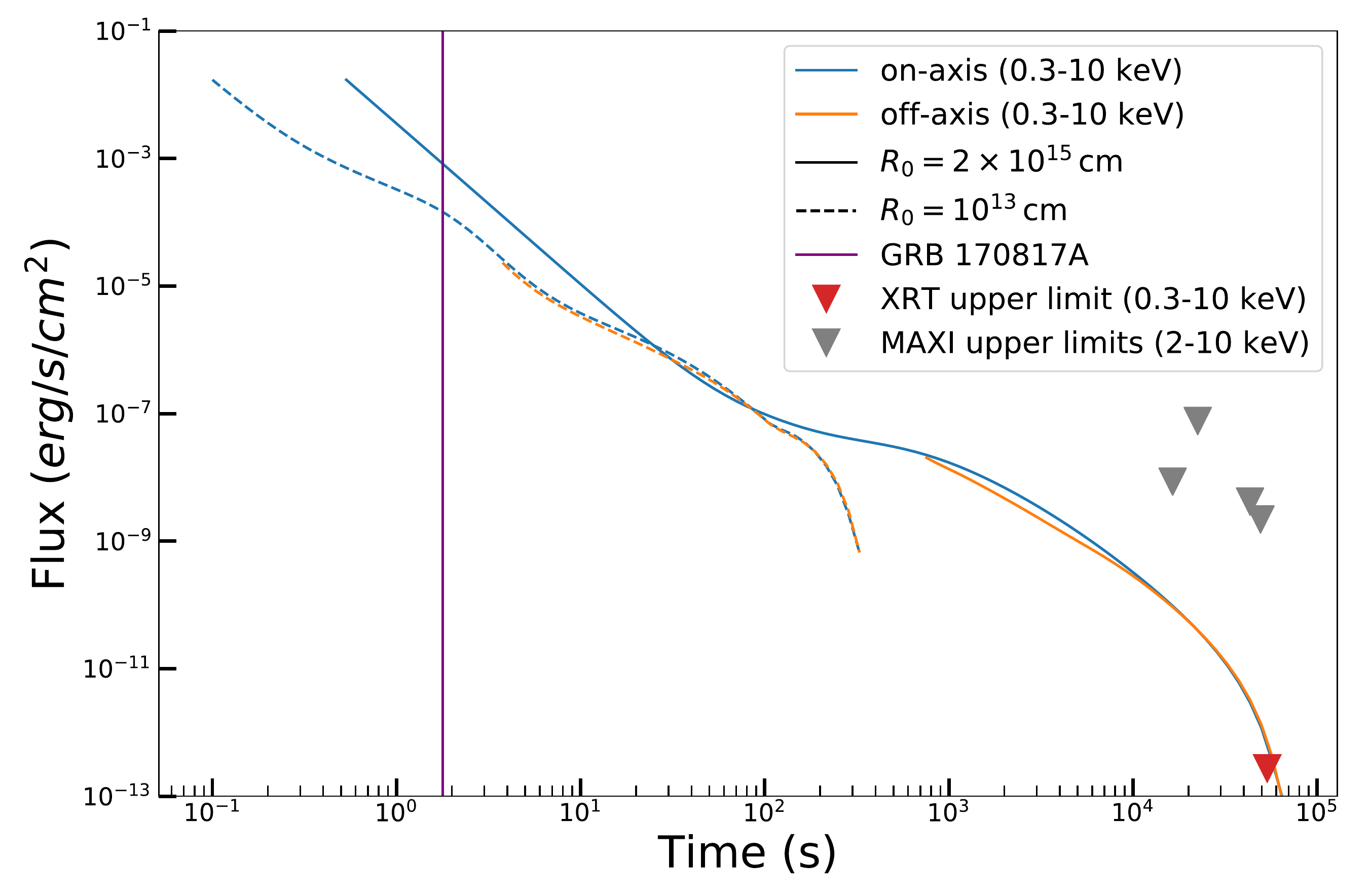}
    \caption{Models of the \ac{HLE} X-ray emission ($0.3-10)\,\rm keV$ of GW170817. Blue and orange line denotes on-axis and off-axis ($\theta_v = 15^\circ$) views and solid and dotted lines marks models with $R_0 = 2\times10^{15}\,\rm cm$ and $R_0 = 10^{13}\,\rm cm$, respectively. Swift-XRT (0.3- 10) keV and MAXI (2-10) keV upper limits are shown by the grey and red triangles, respectively. The purple vertical line represents the trigger time of GRB 170817A with respect to the GW signal. 
    A power-law structured jet is assumed (from \citealt{Ghirlanda2018}) with  $\Gamma_c = 251$, $\theta_c = \theta_E = 3.4^\circ$, $E_{\rm K, ISO} = 2.5\times 10^{52}\, \rm erg$, and  power-law slope $k_c = 3.5$ and $k_E = 5.5$ for the energy and bulk Lorentz factor, respectively. A \ac{SBPL} with  $h\nu_{\rm peak} = 100\,\rm keV$, $\alpha_s = 0.5$ and $\beta_s = 1.3$ is assumed.}
    \label{fig:170817}
\end{figure}

Fig. \ref{fig:170817} shows the \ac{HLE} emission lightcurve  (i.e. flux integrated in the $(0.3-10)\,\rm keV$ energy range) for $R_0 = 2\times 10^{15}\,\rm cm$ and $R_0 = 10^{13}\,\rm cm$ (solid and dashed lines, respectively). The orange lines correspond to the jet seen off-axis (as it should be based on the \ac{EM} modelling of \citealt{Ghirlanda2018}). 
The normalization of the off-axis lightcurve is calculated with respect to the on-axis one, which in turn has been normalized to the luminosity $L_{\rm peak} = 1.3\times 10^{51}\, \rm erg/s$. This value has been obtained assuming central engine duration $\Delta T_{\rm engine} = 2\,\rm s$ and calculating the on-axis isotropic equivalent energy with Eq. 16 of \citet{Salafia2019} with a radiative efficiency $\eta=0.25$ and integrating over the energy range $(0.3-10) \, \rm keV$. 

\emph{Swift}-XRT (0.3-10) keV and MAXI (2-10) keV upper limits are shown by the red and grey triangles, respectively.

From Fig. \ref{fig:170817} we can appreciate that XRT would have been able to observe the \ac{HLE} from the GRB 170817A: the earliest follow up would grant a detection even for small emission radii (dashed lines in Fig. 
\ref{fig:170817}). 
 
With the current follow up timing, the emission, in the case of $R_0 = 2\times10^{15}\,\rm cm$ (solid lines in Fig. \ref{fig:170817}), appears to be marginally detectable. Uncertainty on the parameters of the jet structure are not included in this analysis. 
We conclude that  the current  X-ray follow-up observations of GW170817, starting at relatively late times with respect to the typical timescales of development of the \ac{HLE}  emission, cannot exclude the presence of the latter component while they constrain the emission radius to be  $\lesssim 2 \times 10^{15}\,\rm cm$. 

A more stringent upper limit on the emission radius could be derived under the assumption that the observed $\gamma$--ray emission of GRB 170817A is produced by a standard GRB jet seen off-axis. In this case the first photons of the HLE emission should be coincident with the time of arrival of the $\gamma$--ray photons. The start time of the HLE emission strongly depends on the emission radius (see Fig. \ref{fig:170817}). Within the uncertainties of our model we infer that a considerably small emission radius  $\lesssim 5 \times 10^{12}$ cm is required in order to  produce an HLE emission starting at the  trigger time of GRB 170817A. This result seems to support alternative interpretations, e.g. that the $\gamma$--ray emission of GRB 170817A was likely produced by e.g. the cocoon shock breakout \citep{Kasliwal2017, Gottlieb2018, Nakar2018}.

\section{Summary and Discussion}
\label{sec:summary}

In this work we expand the idea proposed in O20 that interprets the steep decay and the plateau, often observed in \ac{GRB} X-ray afterglows, as the \ac{HLE} of the prompt emission produced by a structured jet. In this work we explore more realistic physical conditions: the finite duration of the emission (as compared to an infinitesimal duration considered in O20) and the possible contribution of a non negligible latitude-dependent optical depth. The main development, with respect to O20, is to consider the HLE as observed from any arbitrary viewing angle. 

Our interpretation of the steep decay-plateau emission as HLE from a structured jet overcomes some of the difficulties of interpreting it as due to a long--lived millisecond magnetar whose ability to launch a jet is uncertain, (\citealt{MurguiaMontes2014, MurguiaMontes2017, CiolfiKastaun2017, CiolfiKastaun2019, Ciolfi2020}; see however \citet{Mosta2020} who pointed out that accounting for neutrino transport in simulations may help the formation of a jet). 

We showed that considering an emission of finite duration (Fig. \ref{fig:finite}) the main conclusions of O20 are not changed: a plateau following the steep decay phase appears the closer is the observer line of sight to the jet axis. It is worth noting that in our framework we considered an entire surface switching on at the same time, expanding while radiating and then switching off at the same time. 
A more realistic situation could be that of a surface that switches on (and off) gradually in time from the axis to the wings, such that the core radiates earlier, when $R_0$ is small, and the wings radiates later, when $R_0$ is large. Even in this case however, a steep decay-plateau behavior is expected, with a steep decay generated by the core at smaller radii and a plateau generated by the wings at larger radii.

The steep decay-plateau structure is preserved even when the effect of Thomson opacity in the emitting region is considered. In this case, however, in order to avoid a considerable suppression of the plateau emission (which being produced by the wings of the jet would suffer mostly from this effect), the emission should take place at large radii (\emph{e.g} $R_0 \gtrsim  10^{15}\,\rm cm$). This condition agrees with the results of O20, namely that a large $R_0$ is required in order to produce a plateau phase lasting more than $\approx 10^{3}$ s\, as often observed\footnote{A large radius is required also from spectral considerations when electron-synchrotron emission is considered (see \emph{e.g.} \citealt{Beniamini2018, Ghisellini2020})}. Interestingly, a small emission radius would produce only a steep decay since the plateau would be absorbed. 

In the context of a finite duration emission,  accounting for the opacity does not impact considerably the lightcurve, since the opacity at earlier times (i.e. smaller $R_0$) affects only the emission at high latitudes, which do not contribute substantially to the total lightcurve  (see Fig. \ref{fig:finite_with_opacity}). Therefore, in the present work we have shown that even when more realistic physical condition are considered, O20's results still hold. As pointed out in O20, with a reasonable choice of parameters, plateau duration up to a few $10^4\, \rm s$ can be recovered. 
Although this timescale can in principle account for the duration of the majority of the events, many of them could be characterized by plateaus lasting longer. One example is GRB060729, which is characterized by a $\sim 4\times 10^4\rm \, s$ long plateau and a final power-law decaying tail observed up to $125\rm \,days$ after the $\gamma$-ray trigger \citep{Grupe2007}. While the duration of the plateau is compatible (even thought marginally) with the results in O20, the 125 days long tail cannot be due to the \ac{HLE} and it likely results from the forward shock emission which is expected to dominate at late times. It is important to stress here that the \ac{HLE} model presented in O20 and in the present work does not pretend to claim that the \ac{HLE} is the only component determining the X-ray lightcurve, but rather that it may give a non negligible (if not dominant) contribution during the plateau phase. 

We note that in the context of the \ac{HLE} model an important aspect that has not been addressed in the present work is the expected spectral evolution of the source. Several studies reported a softening during the steep decay, which terminates at the onset of the plateau, where an hardening can eventually occur \citep{BBZhang2007, Liang2007}. From a qualitative point of view in our model, which involves a \ac{SBPL} spectrum, the spectral evolution is governed by the evolution of the Doppler factor (see Fig. 2 in O20). Its decrease during the steep-decay is expected to lead to a softening in the spectrum, as already pointed out in previous studies \citep{Zhang2009, Genet2009}. Similarly, the constant (slightly increasing) trend of the Doppler factor during the plateau should lead to no evolution (or a slightly hardening) of the spectrum. These considerations, however, are based on the simplistic assumption of a spectrum that is uniform along the emitting surface and that does not evolve in time. A proper modelling of the spectral evolution along with the inclusion of the forward shock contribution will be explored in a future work. It is worth noticing that recently, \citet{Panaitescu2020}, using a model similar in concept to that in O20, but with a more refined prescription for the spectrum, was successful in fitting the lightcurve of four \acp{GRB} (GRB061121, GRB060607A, GRB061110A and GRB061007) accounting also for their spectral evolution.

We studied how the HLE emission lightcurve varies with the viewing angle  and with the jet structure parameters (Sec. \ref{sec:results}). We found the steep--plateau morphology is a characteristic of small viewing angle (i.e. within the jet core aperture) while for off--core observers the lightcurve appears shallower than the (on--axis) steep decay and steeper than the (on--axis) plateau. 

A possible consequence of our model is that  \acp{GRB} with steep decay and plateau emission phases should be those preferentially observed within the jet core. GRBs observed off--core should present less prominent steep--plateaus and also should be appear (at the same distance) fainter. However, for instance, GRB130427A  is one of the most energetic \acp{GRB} but without a clear steep decay --plateau lightcurve \citep{VonKienlin2013, MaselliBeardmore2013, MaselliMelandri2014}.  
However, while a detailed study is out of the scope of the present work, we notice that parameters like $R_0$ and the opacity can in some configuration suppress the plateau component, or the \ac{HLE} component can be outshined by a brighter forward shock emission. 

Concerning the onset of the off-axis emission, our results show that, fixing the structure, depending on the viewing angle we expect the first photon to reach the observer with different delays with respect to the jet launching time. In the context of \acp{GW}, this is an important information both for the \ac{EM} follow-up of \ac{GW} events and 
for the search of \ac{GW} signals triggered by \acp{GRB} \citep{Dietz2013, Kelley2013, Salafia2017, Abbott2019, BurnsGoldstein2019, FermiLVC2020}. This is of particular interest to choose the time window where searching for \acp{GW} in sub-thershold analysis in connection with weak FERMI and \emph{Swift} signals.

Furthermore, we consider that in those cases where the \ac{HLE} emission is not quenched by compactness, its faint off-axis emission can be, in principle, observable by the present X-ray sky surveys such as eROSITA or the future proposed wide field X-ray telescopes such as THESEUS. This is shown in particular in Fig. \ref{fig:FluxAngle}, where, assuming the model of GRB 100906A in O20, we can see that up to $z = 0.045$  (\ac{LVC} range for \ac{BNS} merger) THESEUS and eROSITA are able to detect the \ac{HLE} at large viewing angles.
Therefore, for these instruments the off-axis view of \ac{GRB} prompt emission may result as a new population of X-ray transients.

Finally, we focused on GW170817 and, with the aid of jet parameters derived by \citet{Ghirlanda2018}, we model two possible X-ray \ac{HLE} lightcurves for two limiting emission radii. We found that, provided $R_0 < 2 \times 10^{15}\,\rm cm$, our model predictions are fully consistent with the MAXI and \emph{Swift} XRT observations, since the \ac{HLE} is expected to be too faint for MAXI and would have already faded away before the XRT observation. Furthermore, considering the temporal onset of the off-axis emission measured from the \ac{GW} arrival time, we conclude that in order to explain GRB 170817A as an off-axis prompt emission we would require a radius $R_0 < 5\times 10^{12}\,\rm cm$. This particularly low value of radius, along with the fact that the predicted on-axis maximum flux would have been very faint, suggests us that GRB 170817A is probably not the off-axis view of a standard \ac{GRB}.

\begin{acknowledgements}

We thank Stefan Johannes Grimm for the Einstein Telescope range, Giulia Stratta for the THESEUS-SXI limits and Samuele Ronchini for the discussion about some aspects of the modelling presented in this paper. The authors acknowledges the anonymous referee for her/his suggestions that help us to improve the presentation of our results. SA acknowledges the PRIN-INAF "Towards the SKA and CTA era: discovery, localization and physics of transient sources". MB, SDO, GO acknowledge financial contribution from the agreement ASI-INAF n.2017-14-H.0. MB acknowledge financial support from MIUR (PRIN 2017 grant 20179ZF5KS). SA acknowledges the GRAvitational Wave Inaf TeAm - GRAWITA (P.I. E. Brocato).
GO is thankful to INAF -- Osservatorio Astronomico di Brera for kind hospitality during the completion of this work. This paper is supported by European Union’s Horizon 2020 research and innovation programme under grant agreement No 871158, project AHEAD2020. 

\end{acknowledgements}

\bibliographystyle{aa}  
\bibliography{references} 

\begin{appendix} %First appendix

\section{EATRs equation}
\label{sec:EATReq}

A central aspect of our analysis concerns the identification of the EATRs. In this Appendix we describe the algorithm that we employed to calculate them. 

We consider a general axisymmetric surface described by the function $R = R(\theta)$. This surface corresponds to the emitting surface in our problem. We identify a cartesian coordinate system $\mathcal{K}$ whose z--axis coincides with the surface's symmetry axis. In this system the emitting surface can be described by the following parametrization:
\begin{equation}
    \begin{pmatrix}
        x\\
        y\\
        z
    \end{pmatrix} = R(\theta)
    \begin{pmatrix}
        \sin \theta \cos \phi\\
       \sin \theta \sin \phi\\
        \cos \theta
    \end{pmatrix}.
\end{equation}

We choose the orientation of the $x$ and $y$ axis of this frame such that the observer lies on the $x-z$ plane  (due to the axial symmetry this choice do not imply any loss of generality). 

Now we define a second reference frame $\mathcal{K}'$ obtained applying on $\mathcal{K}$ a rotation around the $y$ axis of an angle $\theta_v$, such that $z'$ coincides with the observer line of sight. The coordinates of a surface point in the new frames are:
\begin{equation}
    \begin{pmatrix}
    x'\\
    y'\\
    z'
    \end{pmatrix} = 
    \begin{pmatrix}
        &\cos\theta_v &0 &-\sin\theta_v\\
        &0 &1 &0\\
        &\sin\theta_v &0 &\cos\theta_v
    \end{pmatrix}
    \begin{pmatrix}
        x\\
        y\\
        z
    \end{pmatrix}.
\end{equation}
In particular we are interested in those surface points that have the same $z'$, because they are those who have the same distance from the observer and consequently those emitting photons with the same arrival time. We consider as reference time the travel time of a photon emitted from the central engine at jet launching time, which corresponds to $t_ 0 = d_L/c$. The travel time of a photon emitted from a point with a given $z'$ is thus $t_{\rm obs} = t_{\rm em} - (d_L - z')/c -t_0$, where $t_{\rm em}$ is the time (measured from $t_0$) at which the emission occurs. Solving for $z'$ we obtain: 
\begin{equation}
    z' = R(\theta)\bigl(\sin\theta_v\sin\theta\cos\phi + \cos\theta_v\cos\theta\bigr) = -c(t_{\rm obs} - t_{\rm em}). 
    \label{eq:EATR}
\end{equation}
This equation is the \emph{EATR equation}, namely its solution, which is obtained numerically, represents a curve $\theta = \theta(\phi)$ on the emitting surface that corresponds to the EATR at that time.

\section{Derivation of the flux}
\subsection{Infinitesimal duration pulse}
\label{sec:AppendixInf}
In this appendix we report the proof of Eq.\ref{eq:flux_PL}.
As in \citet{KumarZhang2015} we start from the general expression of the flux in terms of integral of specific intensity over the observer solid angle:
\begin{equation}
    F_\nu = \int d\Omega_{\rm obs} \cos \theta_{obs} I_\nu.
\end{equation}
It is worth noticing that the factor $\cos \theta_{\rm obs}$, which represents the projection of the emitting surface elementary area along the observer line of sight, is correct only for a spherical surface, so we substitute it with the more general expression $\hat{k}'\cdot \hat{n}$. Moreover, this factor must be present only if the emitting region is optically thick. This is due to the fact that when the matter is optically thin radiation can always reach the observer even if the normal $\hat{n}$ to the emitting surface is orthogonal to the line of sight, because all the photons can freely stream out of the edge of the surface (even if we approximate the emitting surface as a surface the elementary emitters are always contained within a volume, \emph{e.g.} see \citealt{Ghisellini2013}). On the other hand, when the matter is optically thick only a negligible amount of radiation (proportional to the negligible width of the emitting region) will stream out from the surface's edge, and the most of contribution will come from the fraction of the surface "exposed" to the observer. In order to take into account this effect we define the optical depth-dependent projection factor:
\begin{equation}
    P(\theta, \phi, \tau) \equiv |\hat{k}'\cdot \hat{n} (\theta, \phi)| + (1 - |\hat{k}'\cdot \hat{n} (\theta, \phi)|)\times \exp(-\tau),
\end{equation}
where the first term represent the fraction of the emitting surface 'exposed' to the observer and the second therm represents the fraction not exposed to the observer, whose contribution is quenched by the factor $\exp{-\tau}$, where $\tau$ is the optical depth. When the opacity is neglected (as in Sec. \ref{sec:instantaneous}) the projection factor is simply $P(\theta, \phi, \tau = 0) = 1$, while in the optically thick limit (\emph{i.e.} $\tau \gg 1$) $P(\theta, \phi, \tau) \simeq |\hat{k}'\cdot \hat{n} (\theta, \phi)|$. Note that $0 \le P(\theta, \phi, \tau)\le 1$.

The other substitutions we apply involve the solid angle, which is written in terms of elementary area and source distance, $d\Omega_{\rm obs} = d^{-2}_LdS$ and the specific intensity, which is written in terms of specific intensity in the comoving frame. 
Thus we obtain:
\begin{equation}
    F_\nu = \frac{1}{d^2_L(1+z)^3}\int_S dS P(\theta, \phi, \tau) D^{3}I'_{\nu'}(\nu', \theta, t).
    \label{eq:AppendixFlux}
\end{equation}
Now we consider the limiting case in which all the radiation is emitted in an impulse centered at the time $t_{\rm em}$ with infinitesimal duration in the \ac{CoE} frame. We can thus write the intensity in terms of a Dirac delta of time:
\begin{equation}
    I'_{\nu'}(\nu', \theta, t) = \eta'_{\nu'}(\nu', \theta)\delta(t-t_{\rm em}),
    \label{eq:internsity_delta1}
\end{equation}
where $\eta'_{\nu'}(\nu', \theta)$ is the total energy radiated in the comoving frame at frequency $\nu'= \nu (1+z)/D(\theta, \phi)$ by the elementary area at $\theta$, $t$ is the time variable in the \ac{CoE} frame, while $t_{\rm em}$ is a constant corresponding to the actual emission time in which the emission takes place. $\eta'_{\nu'}(\nu', \theta)$ can be further decoupled in two different factors: a frequency-independent part and a spectral part:
\begin{equation}
    \eta'_{\nu'}(\nu', \theta) = \eta'_{\nu'_0}\epsilon(\theta)S'_{\nu'}(\nu', \theta)
    \label{eq:etadefinition}
\end{equation}

where $\epsilon(\theta)$ is the same term appearing in Eq. \ref{eq:structGaussian} (Eq. \ref{eq:structPL}) describing the structure in energy of the emitting surface, $S'_{\nu'}(\nu', \theta)$ represents the spectrum normalized to 1 at $\nu = \nu'_0$ and $\eta'_{\nu'_0}$ is a constant corresponding to $\eta'_{\nu'}(\nu', \theta)$ at $\nu' = \nu'_0$ and $\theta = 0$. It is worth noticing that although the comoving frame spectrum $\eta'_{\nu'_0}S_{\nu'}(\nu')$ is taken as uniform across the emitting surface (hence, it does not depend on $\theta$ and $\phi$), when 
we integrate along the EATR we have to take into account that radiation from different part of the surface are blueshifted by a different factor. Therefore, the integrand in Eq. \ref{eq:AppendixFlux} have to be evaluated in $\nu' = \nu(1+z)/D(\theta, \phi)$, fixing the observer frame frequency $\nu$. This introduce in the $\eta'_{\nu'}$ a further $\theta$ and $\phi$ dependence (through $\nu'$), which preserves the shape of the spectrum and shifts it by a $D(\theta, \phi)$ factor along the $\nu$-axis. 

At this point, we can write $t = D\Gamma t_{\rm obs}$ (see Eq. 11 in \citet{KumarZhang2015}) to make explicit the dependence of the observation time $t_{\rm obs}$, such that Eq.~\ref{eq:internsity_delta1} writes as
\begin{align}
     I'_{\nu'}(\nu', \theta, t_{\rm obs}) &= \eta'_{\nu'}(\nu', \theta)\delta[D(\theta, \phi)\Gamma(\theta)t_{\rm obs}-t_{\rm em}]=\notag\\ &= \eta'_{\nu'}(\nu')\delta[h(\theta, \phi)].
\end{align}
It is worth noticing that the solution of the equation $h(\theta, \phi) = 0$ identifies a curve $\theta = \theta(\phi)$ on the emitting surface, which represents the \ac{EATR} at a given $t_{\rm obs}$.

Now, taking into account the last substitution, we can write the flux density as 
\begin{equation}
    F_\nu(t_{\rm obs}) = \frac{1}{d^2_L(1+z)^3}\int_S dS P(\theta, \phi, \tau) \eta'_{\nu'}(\nu')D^{3}(\theta, \phi)\delta[h(\theta, \phi)].
\end{equation}

Now the Dirac delta allows us to transform the surface integral into a line integral on the EATR. Due to the property of the delta of a multivariate function \citep{Hormander2015analysis} we finally obtain
\begin{equation}
        F_\nu (t_{obs}) = \frac{1}{d^2_L(1+z)^3}\oint_{EATR} \frac{\eta'_{\nu'}(\nu', \theta)D^{3}(\theta, \phi)P(\theta, \phi, \tau)dl}{|\nabla{h}(\theta, \phi)|}. 
\end{equation}

Given this equation, an interesting case worth to be studied is that of a power-law spectrum, that we can describe as:
\begin{equation}
    \eta'_{\nu'}[\nu(1+z)/D, \theta] = \eta'_{\nu'_0}\epsilon(\theta)\Bigl(\frac{\nu'_0}{\nu}\Bigr)^{\beta_s}\frac{D^{\beta_s}(\theta, \phi)}{(1+z)^{\beta_s}},
\end{equation}
where $\beta_s$ is the power-law spectral index. With this spectrum the equation for the flux writes as:
\begin{equation}
        F_\nu (t_{obs}) = \frac{\eta'_{\nu'_0}}{d^2_L}\Bigl(\frac{\nu'_0}{\nu}\Bigr)^{\beta_s}\oint_{EATR} \frac{\epsilon(\theta)D^{3+\beta_s}(\theta, \phi)P(\theta, \phi, \tau)dl}{(1+z)^{3+\beta_s}|\nabla{h}(\theta, \phi)|}. 
\end{equation}
We can test this equation on the limiting case of a spherical ejecta. In this case we set $\theta_v = 0$, $\epsilon(\theta)=1$ and we have also that $\partial_\theta(...) = 0$. In this case the EATRs correspond to circles with length $L=2\pi R \sin \theta$, while the projection factor at a given point of the surface is $P(\theta, \phi, \tau = 0) = 1$  (or $ P(\theta, \phi, \tau) = |\hat{k}' \cdot \hat{n}(\theta, \phi)| = \cos \theta$ in the optically thick regime) and the gradient is $|\nabla h(\theta, \phi)| = D\Gamma \sin \theta/c$.
Considering the previous substitution we obtain:
\begin{equation}
    F_\nu(t_{\rm obs}) = \frac{2\pi\eta'_{\nu'_0}Rc}{d^2_L[\Gamma(1+z)]^{3+\beta_s}}\Bigl(\frac{\nu'_o}{\nu}\Bigr)^{\beta_s}t^{2+\beta_s}_{\rm em}t^{-(2+\beta_s)}_{\rm obs}
\end{equation}
which is consistent with the classical result of \citet{Kumar2000}.

We consider also the flux described by the \ac{SBPL} function in Eq. \ref{eq:SBPL}. 
In this case we have:

\begin{equation}
    \eta_\nu(\nu, \theta, \phi) =  \frac{2\eta'_{\nu'_0}\epsilon(\theta) D^3(1+z)^{-3}}{\Bigl(\frac{\nu}{\nu'_0}\Bigr)^{\alpha_s}(1+z)^{\alpha_s}D^{-\alpha_s} + \Bigl(\frac{\nu}{\nu'_0}\Bigr)^{\beta_s}(1+z)^{\beta_s}D^{-\beta_s}}
\end{equation}
where $\alpha_s$ and $\beta_s$ are the spectral indices below and above the peak. Here we hide the $\theta$ and $\phi$ dependence of the Doppler factor for the seek of compactness. With this spectrum profile the flux writes as: 

\begin{align}
     F_\nu (t_{obs}) = & \frac{2\eta'_{\nu'_0}}{d^2_L(1+z)^3}  \times\notag\\&
     \oint_{EATR} \frac{\epsilon(\theta)D^{3}(\theta, \phi)P(\theta, \phi, \tau)dl}{\Bigl[\Bigl(\frac{\nu(1+z)}{\nu'_0D}\Bigr)^{\alpha_s} + \Bigl(\frac{\nu(1+z)}{\nu'_0D}\Bigr)^{\beta_s}\Bigr]|\nabla{h}(\theta, \phi)|},
    \label{eq:flux_Band}
\end{align}

\subsection{Finite duration pulse}
\label{sec:AppendixFinite}
To account for a finite duration pulse we adopted the following approximate approach: 

\begin{itemize}
    \item we define a specific intensity temporal profile $I'_{\nu'_0}(t)$. The total emitted energy per unit area, per unit frequency and per unit solid angle writes as:
    \begin{equation}
        \eta'_{\nu'_0} = \int^\infty_{-\infty}I'_{\nu'_0}(t)dt.
    \end{equation}
    
    \item We divide the flux profile in $N$ time intervals of finite length $\Delta t_{\rm em}$ and central values $t_{\rm em, i}$ with $i = 1, ..., N$. The energy per unit area radiated in the i-th time interval is:
    \begin{equation}
        \eta'_{\nu'_0, i} \equiv \int^{t_{\rm em, i} + \Delta t_{\rm em}/2}_{t_{\rm em, i} - \Delta t_{\rm em}/2} I'_{\nu'_0}(t)dt.
        \label{eq:norm_finite}
    \end{equation}
    By definition we have that:
    \begin{equation}
        \eta'_{\nu'_0} = \sum_i\eta'_{\nu'_0,i}
        \label{eq:energy_conservation}
    \end{equation}
    
    \item We introduce a function $\Tilde{I}'_{\nu'_0}(t)$ which approximates the flux profile and is defined as a Dirac comb:
    \begin{equation}
        \Tilde{I}'_{\nu'_0}(t) \equiv \sum_i \eta'_{\nu'_0, i}\Bigl(\frac{R_0}{R_0 +\Delta R_{0,i}}\Bigr)^2 \delta(t- t_{\rm em, i}) 
        \label{eq:finiteemissionintensity}
   \end{equation}
   here $\Delta R_{0,i} \equiv R_{0,i}-R_0$ is the difference between the radius of the first and the i-th emitting surface at $\theta = 0$. The term in parenthesis have been added since in the case of an internal dissipation mechanism, as we are considering here, the number of emitters are expected to conserve and dilute in time while the emitting surface expands. With the prescription in Eq. \ref{eq:finiteemissionintensity} the specific intensity approximates as $I_\nu(t, \theta, \phi) = \tilde{I}'_{\nu'_0}(t)D^3(\theta, \phi)\epsilon(\theta)S_\nu(\nu)$.
   
   \item We express the final flux as the superimposition of the N impulses of the Dirac comb:
   \begin{align}
       F_\nu = & \frac{1}{d^2_L(1+z)^3}\int_SdS P(\theta, \phi, \tau) D^{3}(\theta, \phi)\Tilde{I}'_{\nu'_0}(t)\epsilon(\theta)\,\times\notag \\ &S'_{\nu'}[\nu (1+z)/D(\theta, \phi)].
   \end{align}
   Here $S = S_1 \cup ... \cup S_N$, where $S_i$ is the emitting surface of the i-th delta of the comb.
\end{itemize}

It is worth noting that in this approach Eq. \ref{eq:energy_conservation} guarantees that the total energy of the pulse is preserved regardless of the value of N.

\section{Optical Depth Derivation}
\label{sec:AppendixOpacity}

In this Section we derive Eq.~\ref{eq:optical_depth}. We consider a shell of material of inner radius $R(\theta)$ and width $\Delta R(\theta)$ and a photon emitted at $R$. We limit ourselves to photons travelling radially outwards, as the emission is always dominated by them. The optical depth of the shell due to Thomson scattering will be
\begin{equation}
    d\tau(\theta) = \sigma_T n_e(r, \theta) dR(\theta),
    \label{eq:diff_optical_depth1}
\end{equation}
where $n_e(r, \theta)$ is the electron number density and $dR(\theta)$ is the length travelled by photon within the shell. Since the shell is moving at the velocity $\beta c$ in the same time in which the photon travels a distance $dr = c dt$ the shell travels a distance $dr_s(\theta) = \beta(\theta) c dt$, such that $dR(\theta) = dr - dr_s = (1 - \beta)cdt$ \citep{Abramowicz1991}. In the meanwhile, the jet expansion causes matter to dilute, with the electron density decreasing as $n_e(r, \theta) = n_e(R, \theta)(r/R)^{-2}$.

The value of $n_e(R, \theta)$ can be obtained considering that in a time interval $\Delta t$ the amount of isotropic equivalent mass of charged particle passing through $r_0$ is: 
\begin{equation}
    M_{\rm ISO}(\theta) = 4\pi n_e(R, \theta) m_p R^2 \beta c \Delta t = \frac{Y_e L_{\rm K, ISO}(\theta)\Delta t}{(\Gamma-1) c},
\end{equation}
where $Y_e$ is the electron fraction and $L_{\rm K, ISO}$ is the isotropic equivalent kinetic luminosity defined as $L_{\rm K, ISO}(\theta) = (\Gamma(\theta)-1) M(\theta) c^2/\Delta t$. Solving for $n_e(R, \theta)$ we obtain:
\begin{equation}
    n_e(R) = \frac{Y_e L_{\rm K, ISO}}{4\pi R^2 m_p \beta c^3 (\Gamma-1)}.
\end{equation}
Substituting the density profile in Eq. \ref{eq:diff_optical_depth1} we obtain:
\begin{equation}
    d\tau(\theta)  = \frac{Y_e \sigma_T L_{\rm K, ISO}}{4\pi m_p c^3\Gamma^2(\Gamma -1)(1+\beta)r^2}
    \label{eq:diff_optical_depth2}
\end{equation}
The optical depth is then obtained integrating Eq. \ref{eq:diff_optical_depth2}:
\begin{equation}
    \tau(\theta)= \frac{Y_e \sigma_T L_{\rm K, ISO}}{4\pi m_p c^3\Gamma^2(\Gamma -1)(1+\beta)} \int^{R + \Delta R/(1-\beta)}_{R}\frac{dr}{r^2},
\end{equation}
where $\Delta R/(1-\beta) = \Delta R(1+\beta)\Gamma^2$ is the length traveled by the photon to escape the shell. Solving the integral we obtain: 
\begin{equation}
    \tau(\theta) = \frac{Y_e \sigma_T L_{\rm K, ISO}}{4\pi m_p c^3\Gamma^2(\Gamma -1)(1+\beta)}\Bigl[\frac{1}{R} - \frac{1}{R + \Delta R(1+\beta)\Gamma^2}\Bigr],
\end{equation}
where the width of the shell is determinate by the duration of the central engine as $\Delta R(\theta) = \beta(\theta)c \Delta T_{\rm engine}$.

\section{Power-law structures}
\label{sec:Appendix-PL-structure}
\begin{figure*}
    \centering
    \includegraphics[width = \textwidth]{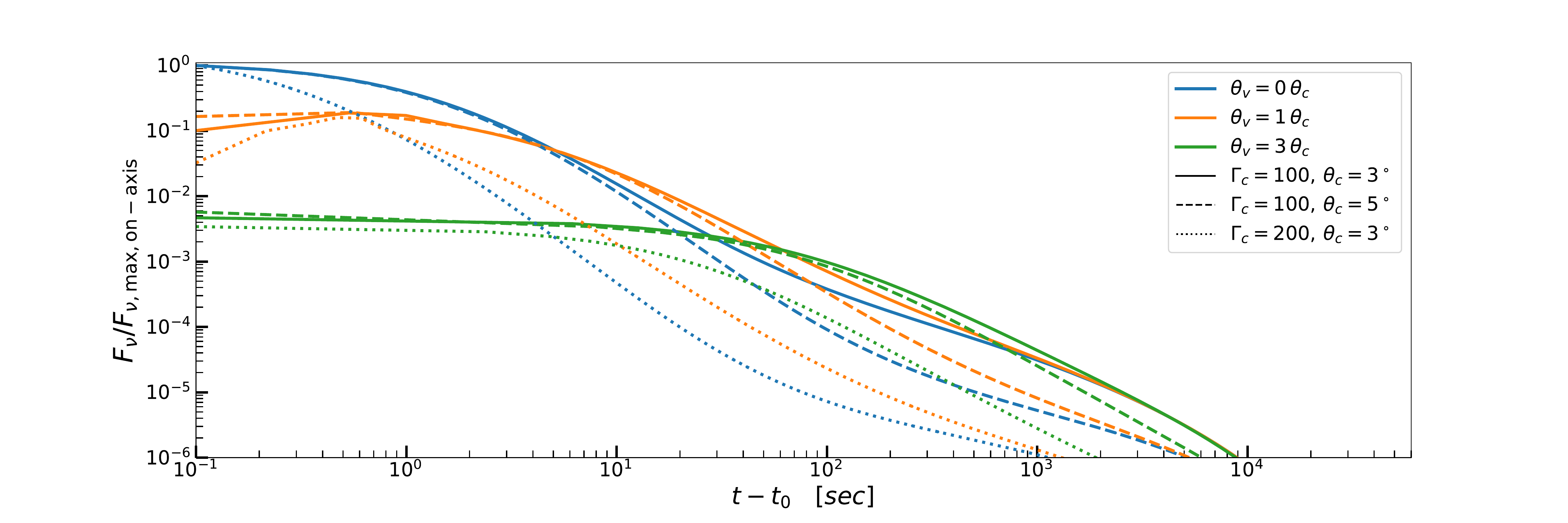}
    \caption{Same as Fig. \ref{fig:case_a} but with a power-law structure with $k=2$.}
    \label{fig:case_c}
\end{figure*}

\begin{figure*}
    \centering
    \includegraphics[width = \textwidth]{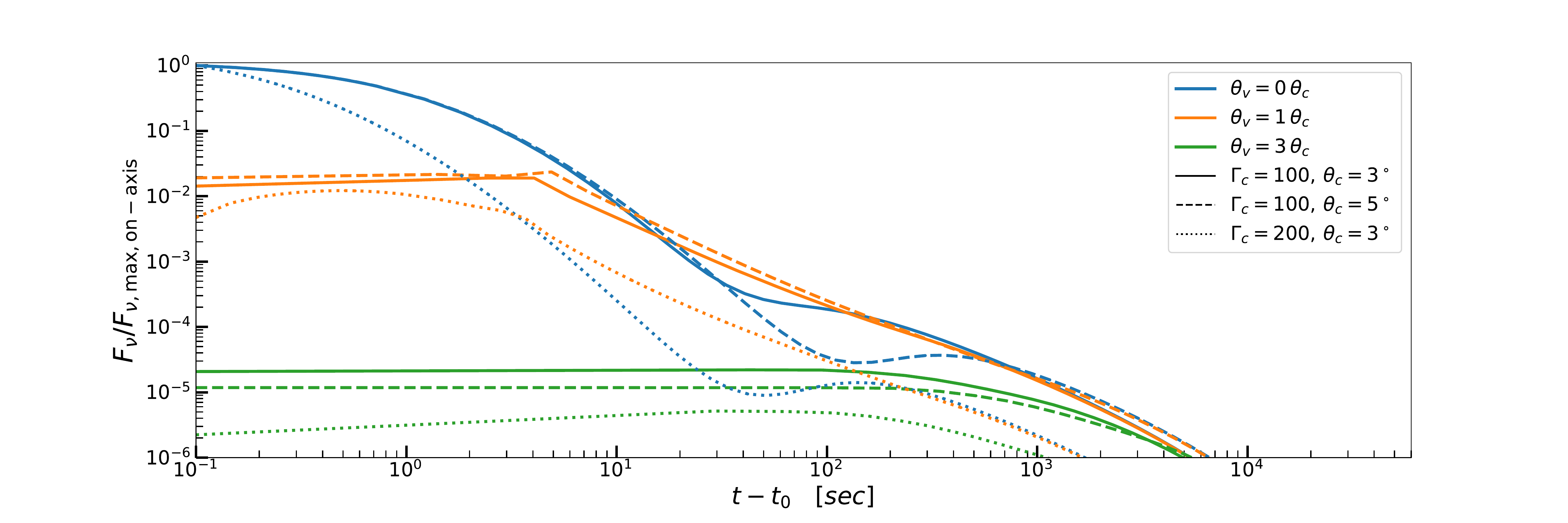}
    \caption{Same as Fig. \ref{fig:case_a} but with a power-law structure with $k=5$.}
    \label{fig:case_d}
\end{figure*}

In this Appendix we provide lightcurves for two different power-law strucures. Finally, Fig. \ref{fig:case_c} and \ref{fig:case_d} show the lightcurves obtained by a power-law structure with $k=2$ and $k=5$, respectively. In both cases the qualitative behavior is the same of the fiducial model. However, the $k=2$ lightcurve presenting more similarities between different viewing angle and a less varying slope in time. As expected, the sharp structure $k=5$ is closer to the Gaussian model.

\end{appendix}

%%%% End of aa.dem
\end{document}